\documentclass[aps,prb,twocolumn,amsmath,amssymb,showpacs,superscriptaddress]{revtex4-1}

\usepackage{graphicx}
\usepackage{amsmath,amssymb,amsfonts}
\usepackage{textcomp}
\usepackage{gensymb}
\usepackage{verbatim}
\usepackage{amsmath}
\begin{document}
\title{Dome of magnetic order inside the nematic phase of sulfur-substituted FeSe under pressure}
%\title{Pressure-temperature-magnetic field phase diagrams of Fe(Se$_{1-x}$S$_{x}$)}
\author{Li Xiang}
\affiliation{Ames Laboratory, Iowa State University, Ames, Iowa 50011, USA}
\affiliation{Department of Physics and Astronomy, Iowa State University, Ames, Iowa 50011, USA}
\email[]{ives@iastate.edu}
\author{Udhara S. Kaluarachchi}
\affiliation{Ames Laboratory, Iowa State University, Ames, Iowa 50011, USA}
\affiliation{Department of Physics and Astronomy, Iowa State University, Ames, Iowa 50011, USA}
\author{Anna E. B\"ohmer}
\affiliation{Ames Laboratory, Iowa State University, Ames, Iowa 50011, USA}
\author{Valentin Taufour}
\affiliation{Ames Laboratory, Iowa State University, Ames, Iowa 50011, USA}
\affiliation{Department of Physics, University of California, Davis, California 95616, USA}
\author{Makariy A. Tanatar}
\affiliation{Ames Laboratory, Iowa State University, Ames, Iowa 50011, USA}
\affiliation{Department of Physics and Astronomy, Iowa State University, Ames, Iowa 50011, USA}
\author{Ruslan Prozorov}
\affiliation{Ames Laboratory, Iowa State University, Ames, Iowa 50011, USA}
\affiliation{Department of Physics and Astronomy, Iowa State University, Ames, Iowa 50011, USA}
\author{Sergey L. Bud'ko}
\affiliation{Ames Laboratory, Iowa State University, Ames, Iowa 50011, USA}
\affiliation{Department of Physics and Astronomy, Iowa State University, Ames, Iowa 50011, USA}
\author{Paul C. Canfield}
\affiliation{Ames Laboratory, Iowa State University, Ames, Iowa 50011, USA}
\affiliation{Department of Physics and Astronomy, Iowa State University, Ames, Iowa 50011, USA}
\email[]{canfield@ameslab.gov}

\date{\today}

\begin{abstract}
The pressure dependence of the structural, magnetic and superconducting transitions and of the superconducting upper critical field were studied in sulfur-substituted Fe(Se$_{1-x}$S$_{x}$). Resistance measurements were performed on single crystals with three substitution levels ($x$=0.043, 0.096, 0.12) under hydrostatic pressures up to 1.8 GPa and in magnetic fields up to 9 T, and compared to data on pure FeSe. Our results illustrate the effects of chemical and physical pressure on Fe(Se$_{1-x}$S$_{x}$). On increasing sulfur content, magnetic order in the low-pressure range is strongly suppressed to a small dome-like region in the phase diagrams. However, $T_s$ is much less suppressed by sulfur substitution and $T_c$ of Fe(Se$_{1-x}$S$_{x}$) exhibits similar non-monotonic pressure dependence with a local maximum and a local minimum present in the low pressure range for all $x$. The local maximum in $T_c$ coincides with the emergence of the magnetic order above $T_c$. At this pressure the slope of the upper critical field decreases abruptly. The minimum of $T_c$ correlates with a broad maximum of the upper critical field slope normalized by $T_c$.

% And $T_s$ of different substitution is suppressed at different rates which indicates that chemical pressure and physical pressure are not equivalent in this system.
  
\end{abstract}

\maketitle 

\section{Introduction}
%Ever since the first discovery of Fe-bases superconductor LaFeAsO Ref [], there has been a worldwide effort to study and expand this superconducting family. To date, various iron-based superconductors have been discovered and categorized into different classes according to there crystal structure, 1111,111,11,122,1144. %Structurally, they all have common layered structure with Fe-As/Se layers stacking together or separated by other elements. 

Despite a large number of different compounds, many iron-based superconductors share similar physical properties. A characteristic feature of this material class is rich phase diagrams, containing an antiferromagnetic phase, which is suppressed upon substitution or pressure, and superconductivity, which emerges at a critical value of this tuning parameter\cite{Canfield2010,Paglione2010}. Usually, the antiferromagnetic ordering is of stripe-type and is preceded or accompanied by a structural tetragonal-to-orthorhombic distortion, associated with electronic nematic order\cite{Fernandes2014}. The magnetic and structural transitions typically extrapolate to zero temperature near the maximum of the superconducting $T_c$ dome, suggesting the possibility that magnetic or nematic fluctuations surrounding a quantum critical point mediate superconductivity\cite{Kasahara2010,Putzke2014}.

Among all the iron-based superconductors, the structurally most simple binary compound, FeSe, does not share this common behavior. First, the structural and magnetic transitions are well separated\cite{McQueen2009}; at ambient pressure the structural transition occurrs at $T_s=90$ K with no signature of magnetic order observed at ambient pressure down to 0.24 K (Ref. \onlinecite{Bendele2010}). Very recent specific heat indicates a possible antiferromagnetic transition at 1.08 K (Ref. \onlinecite{Chen2017}). However, their results contradict previous results\cite{Lin2011}, in which no anomaly in the specific measurement near this temperature was observed.

Second, under approximately 0.8 GPa of applied pressure, magnetic order clearly emerges\cite{Bendele2010,Bendele2012,Terashima2015,Kaluarachchi2016} above $T_c$ and the magnetic transition temperature $T_m$ exhibits a dome-like pressure dependence between 0.8 GPa and 6 GPa\cite{Sun2016a,Sun2017}. Strong coupling between orthorhombic distortion and magnetic order under pressure was demonstrated\cite{Kothapalli2016}. Nevertheless, the large separation of $T_s$ and $T_m$ at ambient pressure raises the question of how the nematic order and magnetism are related in this compound\cite{Glasbrenner2015,Yu2015,Wang2015,Chubukov2016}.  

Third, the pressure dependence of superconducting transition temperature $T_c$ shows a remarkable non-monotonic structure, with a local maximum of $T_c$ around 0.8 GPa, a local minimum around 1.2 GPa, a plateau around 4 GPa and finally a maximum of $37$ K around 6 GPa, before $T_c$ decreases at even higher pressures \cite{Miyoshi2014,Kaluarachchi2016,Sun2016a}. Moreover, recent studies found that monolayer thin films of FeSe on STO shows superconducting behavior at temperatures higher than 100 K\cite{Ge2015}. Hence, FeSe gives us a unique opportunity to study how nematicity, magnetism and superconductivity interact with each other.

The maximum $T_c$ of bulk FeSe under pressure is achieved in the pressure range above 5 GPa. However, FeSe has a complex and interesting phase interplay in the pressure range below 2 GPa. In this pressure range falls the intersection of the nematic phase, magnetic order and superconductivity\cite{Terashima2015,Kaluarachchi2016,Sun2016a}. 
Several studies have investigated the effect of sulfur substitution on FeSe\cite{Mizuguchi2009,Watson2015II,Coldea2016,Hosoi2016,Ovchenkov2016}. Similar to applied pressure, sulfur substitution suppresses $T_s$. In contrast to pressurized FeSe, no magnetic ordering is found in the substitution-temperature phase diagram of Fe(Se$_{1-x}$S$_x$) and $T_c$ is only moderately enhanced to ~11 K by substitution\cite{Watson2015II}. 
%In the parent compound, no feedback of superconductivity on the orthorhombic distortion has been shown, though a small decrease of magnetic order below Tc under pressure has been seen. In S-doped FeSe, uniquely, the orthohrombic distortion even increases below Tc\cite{Wang2016II}. Overall, the relation of nematicity and superconductivity is not clear. 
In this work, we combine chemical pressure through sulfur substitution up to 12\% and physical pressure up to 1.8 GPa and show that the pressure-induced magnetic phase is strongly suppressed upon substitution in this pressure range. In contrast, the nematic phase and superconducting phase are quite robust and
their behaviors under pressure do not change qualitatively.
%their characteristic pressure dependencies do not change under substitution.

\section{Experimental details}
High quality single crystals of FeSe$_{1-x}$S$_{x}$ ($x=0.043(5)$, $x=0.096(1)$, $x=0.12(2)$) with sharp superconducting transitions at ambient pressure (see Fig 2-5 (b) below), were grown using chemical vapor transport, similar to Ref. \onlinecite{Boehmer2016}. The doping ratio $x$ was determined by energy-dispersive x-ray spectroscopy (EDS) and the given values and errors correspond to the average and standard deviation of EDS results obtained on $\sim10$ spots from typically 3 different samples per batch, respectively.  
%Samples were cut and cleaved to ~ 0.5 mm sizes. 
The $c$-axis resistance was measured on samples with doping ratio $x=0.043$ and $0.096$ of approximate dimensions of $(0.5\times0.5\times0.1)$ mm$^3$, using a two-probe technique similar to Refs. \onlinecite{Kaluarachchi2016,Tanatar2009}. Two Ag wires were attached to the samples by soldering with In-Ag alloy\cite{Kaluarachchi2016,Tanatar2016}. The contact resistance is less than 50 $\mu \Omega$ which is much smaller than the sample resistance of approximately 10 m$\Omega$. Four-probe wiring was used down to the sample contacts. 
The in-plane resistance was measured on a sample with doping ratio $x=0.12$ of approximate dimensions of $(1\times0.5\times0.1)$ mm$^3$ in a standard four-contact configuration, with contacts prepared using silver epoxy. AC resistance measurement were performed in a Quantum Design Physical Property Measurement System using 1 mA; 17 Hz excitation, on cooling and warming at a rate of 0.25 K/min. A Be-Cu/Ni-Cr-Al hybrid piston-cylinder cell similar to the one described in Ref. \onlinecite{Budko1984} was used to apply pressure. Pressure values at low temperature were inferred from the $T_{c}(p)$ of lead\cite{Bireckoven1988}. Good hydrostatic conditions were achieved by using a 4:6 mixture of light mineral oil:n-pentane as pressure medium, which solidifies at room temperature in the range $3-4$ GPa, i.e., well above our maximum pressure\cite{Budko1984,Kim2011,Torikachvili2015}. %The orientation of the samples in the pressure cell were adjusted so that the magnetic field was applied parallel to the $c$-axis. The measurement configuration with parallel current and magnetic field, $j||H||c$ minimizes the contribution of flux flow to the superconducting transitions, which is expected to facilitate the determination of the superconducting upper critical field $H_{c2,c}$.

\section{Pressure-temperature phase diagrams}% of Fe(Se$_{1-x}$S$_x$)}

\begin{figure}
	\includegraphics[width=8.6cm]{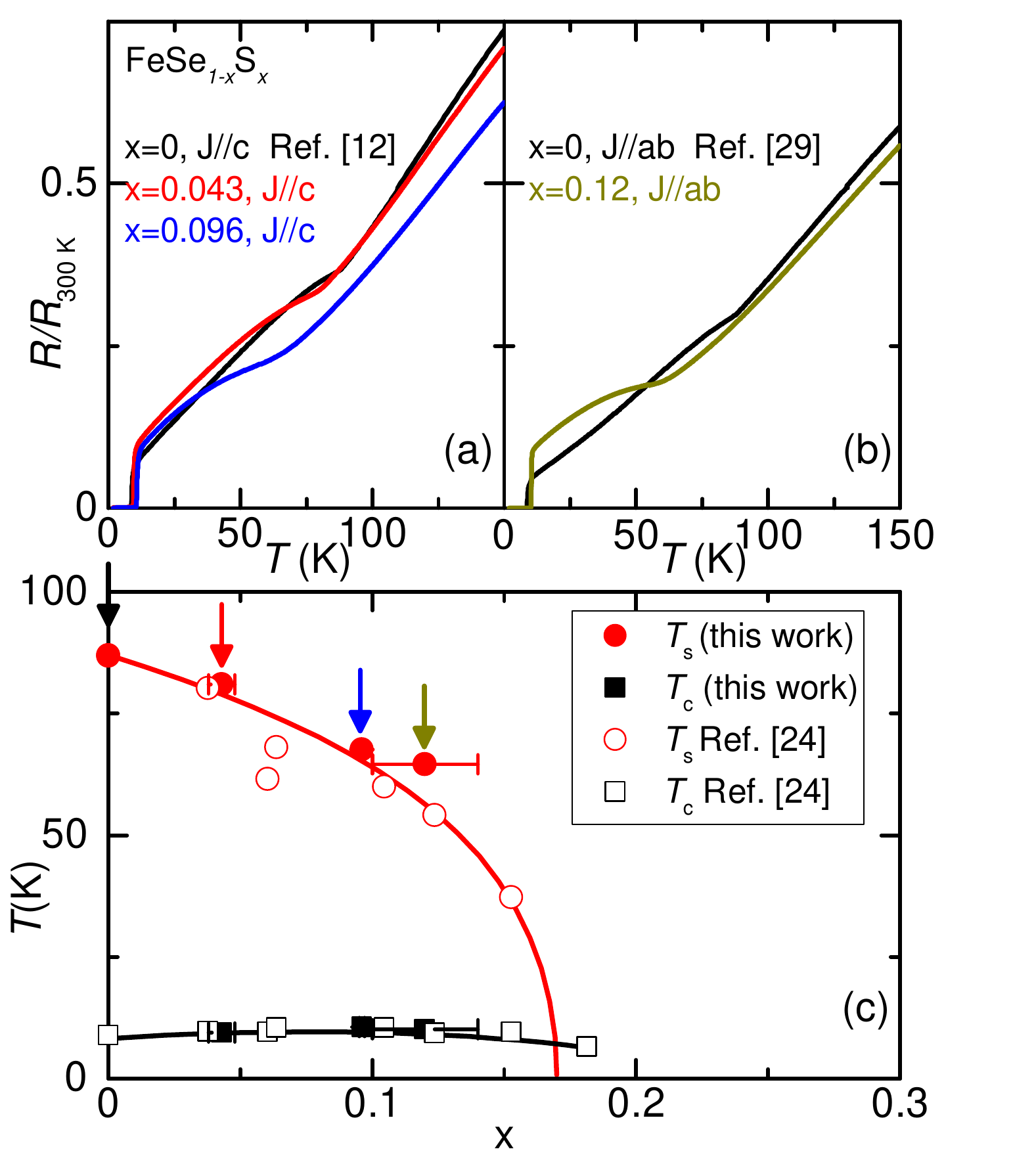}%
	\caption{(a) Temperature dependence of the normalized resistance of Fe(Se$_{1-x}$S$_x$) single crystals with current applied along $c$-axis for $x=0$, $x=0.043$ and $x=0.096$. The data on the parent compound FeSe are taken from Ref. \onlinecite{Kaluarachchi2016}. (b) Temperature dependence of the normalized resistance of Fe(Se$_{1-x}$S$_x$) single crystals with current applied in the ab plane for $x=0$ and $x=0.12$. The data on the parent compound FeSe are taken from Ref. \onlinecite{Tanatar2016}. (c) Doping-temperature phase diagram of Fe(Se$_{1-x}$S$_x$). The four compounds we used in this work are marked. Open symbols are data taken from Ref. \onlinecite{Coldea2016}.
		\label{raw0}}
\end{figure}

Figure\,\ref{raw0} shows the ambient-pressure resistance of the studied Fe(Se$_{1-x}$S$_x$) samples. The resistance is normalized at 300 K. The $c$-axis and in-plane resistance data on the parent compound FeSe are taken from Ref. \onlinecite{Kaluarachchi2016} and \onlinecite{Tanatar2016} respectively. $T_c$ increases slightly from 8.9 K for undoped FeSe to 10.1 K for $x=0.12$. The structural transition, visible as a kink in the resistance data, is suppressed from 90 K to 60 K at the highest studied substitution level. Note that in this work, the in-plane resistance is studied for the $x=0.12$ sample, but $c$-axis resistance for the other three substitution levels. The features at $T_s$ in in-plane and inter-plane resistance are rather similar. The positions of the studied compositions are indicated in the composition-temperature phase diagram in Fig.\,\ref{raw0}(c).

\begin{figure}
	\includegraphics[width=8.6cm]{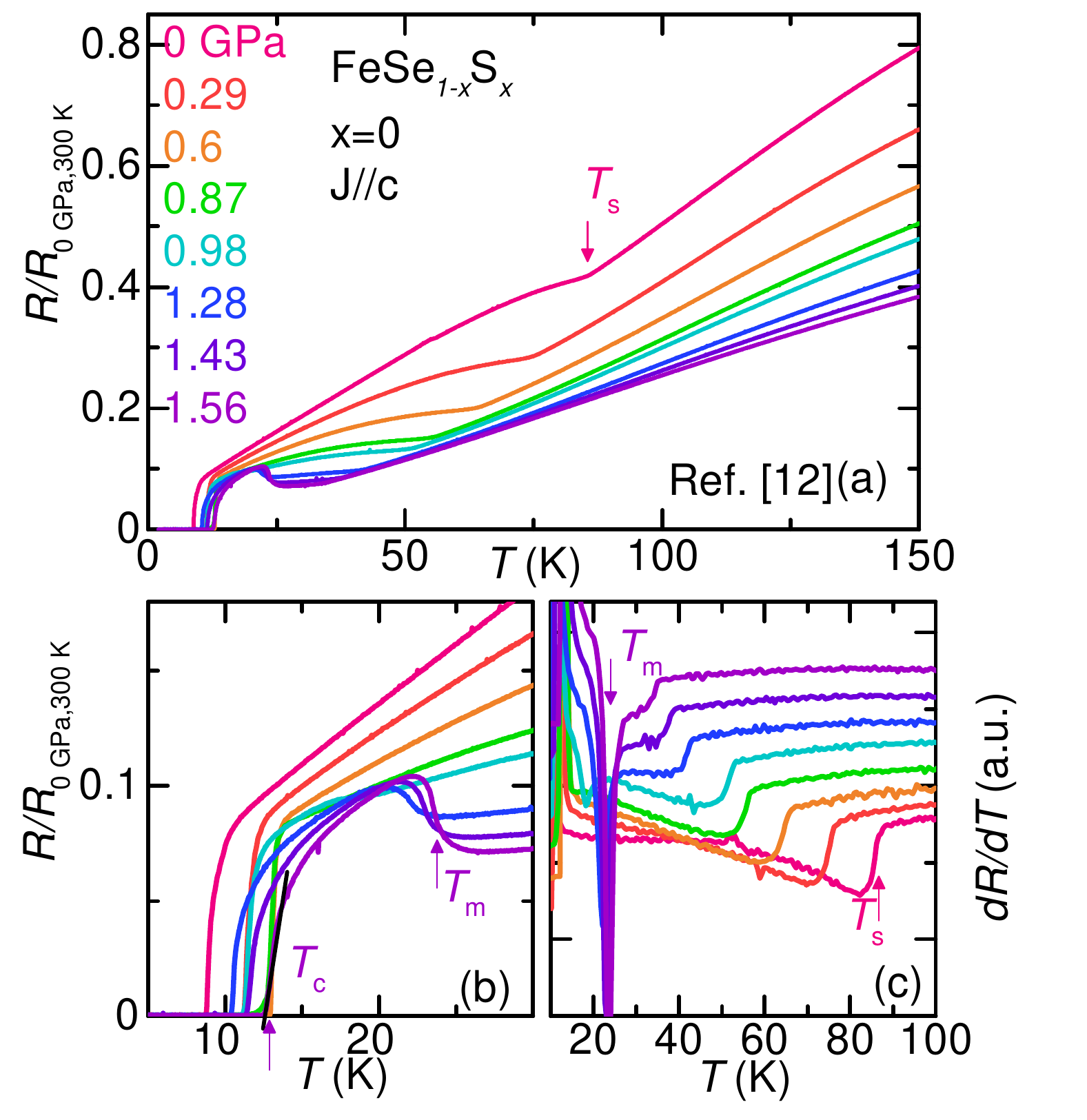}%
	\caption{(a) Evolution of the $c$-axis resistance with hydrostatic pressure for pure FeSe. Data were normalized at room temperature, ambient pressure. (b) Blow up of the low-temperature region. (c) Temperature derivative $dR$/$dT$ showing the evolution of structural transition $T_s$. Data are taken from Ref. \onlinecite{Kaluarachchi2016}. Examples of transition temperatures $T_s$, $T_m$ and $T_c$ are indicated by arrows.
		\label{raw4}}
\end{figure}

\begin{figure}
	\includegraphics[width=8.6cm]{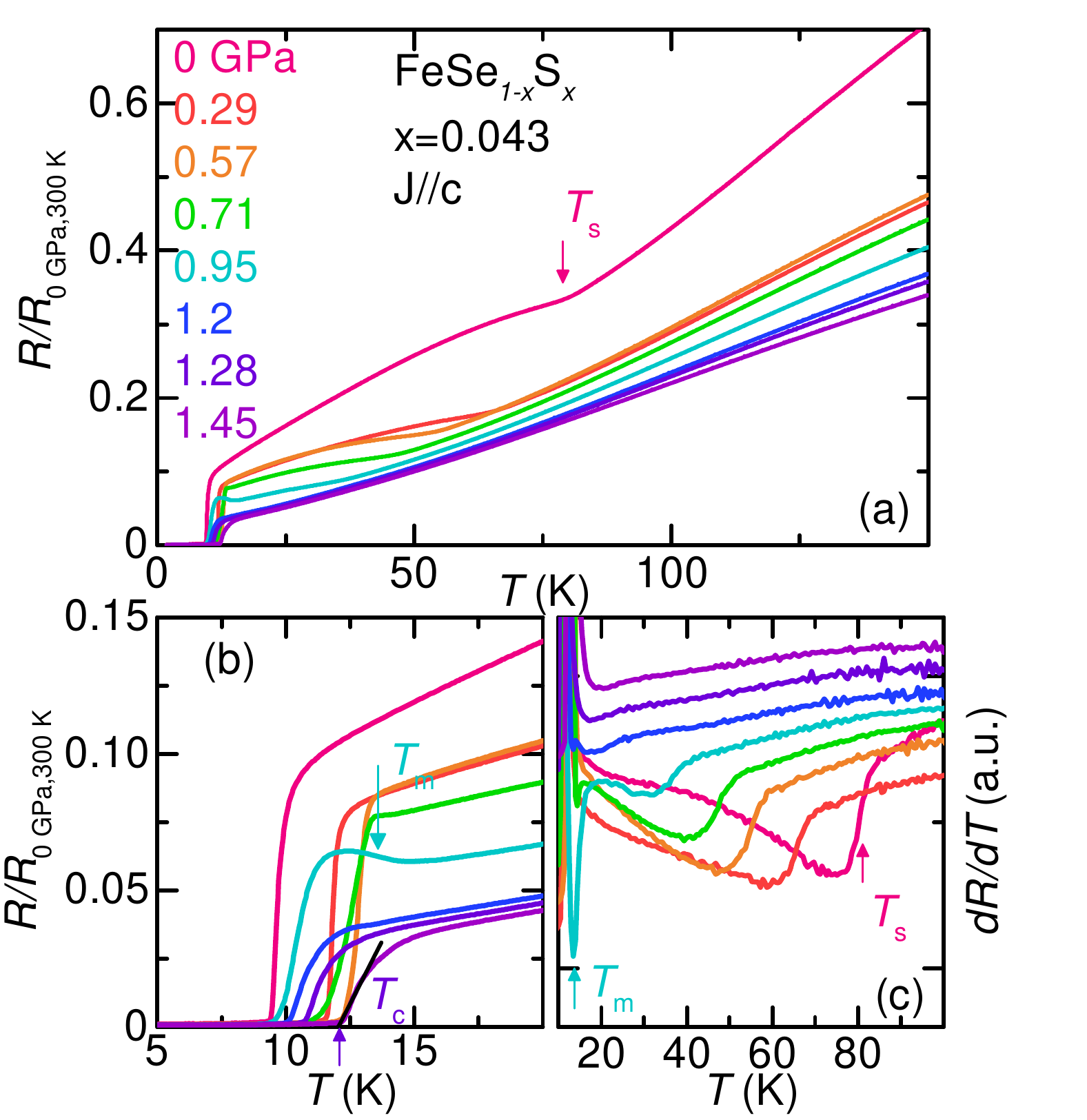}%
	\caption{(a) Evolution of the $c$-axis resistance with hydrostatic pressure for Fe(Se$_{1-x}$S$_x$), $x=0.043$. Data were normalized at room temperature, ambient pressure. (b) Blow up of the low-temperature region. (c) Temperature derivative $dR$/$dT$ showing the evolution of structural transition $T_s$. Examples of transition temperatures $T_s$, $T_m$ and $T_c$ are indicated by arrows. 
		\label{raw1}}
\end{figure}
\begin{figure}
	\includegraphics[width=8.6cm]{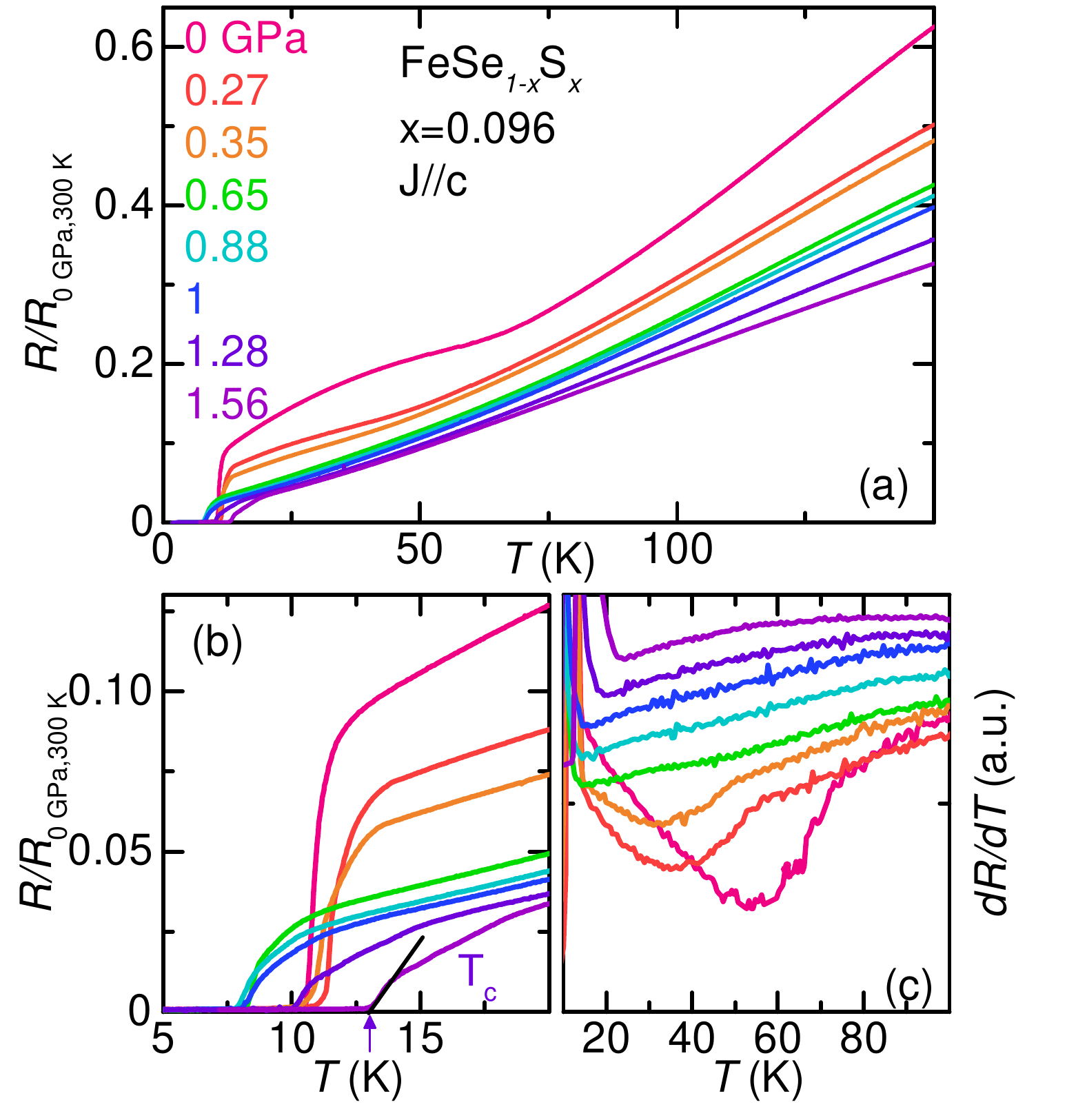}%
	\caption{(a) Evolution of the $c$-axis resistance with hydrostatic pressure for Fe(Se$_{1-x}$S$_x$), $x=0.096$. Data were normalized at room temperature, ambient pressure. (b) Blow up of the low-temperature region. (c) Temperature derivative $dR$/$dT$ showing the evolution of structural transition $T_s$. 
		\label{raw2}}
\end{figure}
\begin{figure}
	\includegraphics[width=8.6cm]{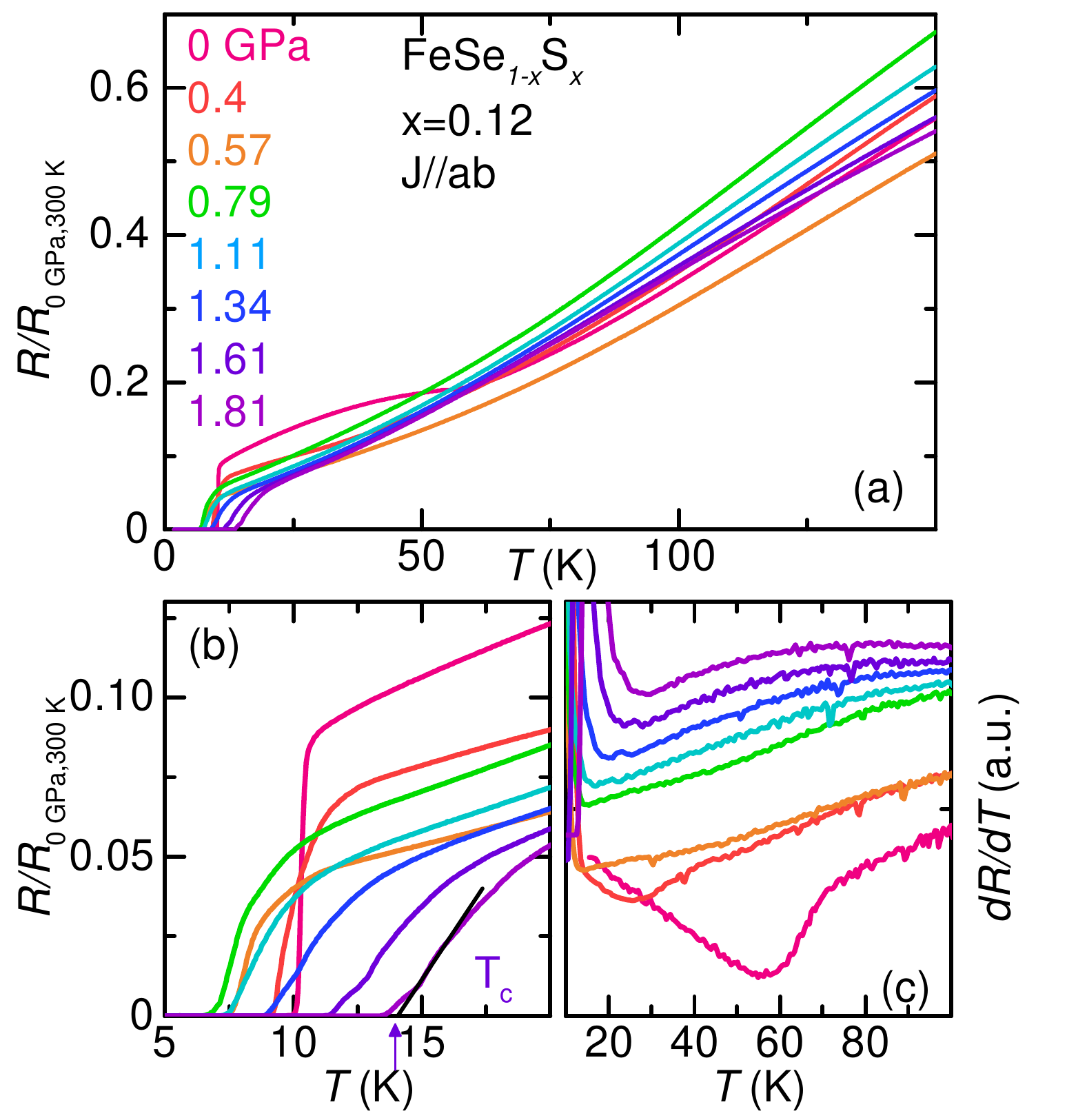}%
	\caption{(a) Evolution of the $c$-axis resistance with hydrostatic pressure for Fe(Se$_{1-x}$S$_x$), $x=0.12$, with in-plane current. Data were normalized at room temperature, ambient pressure. (b) Blow up of the low-temperature region. (c) Temperature derivative $dR$/$dT$ showing the evolution of structural transition $T_s$.
		\label{raw3}}
\end{figure}
\begin{figure}
	\includegraphics[width=8.6cm]{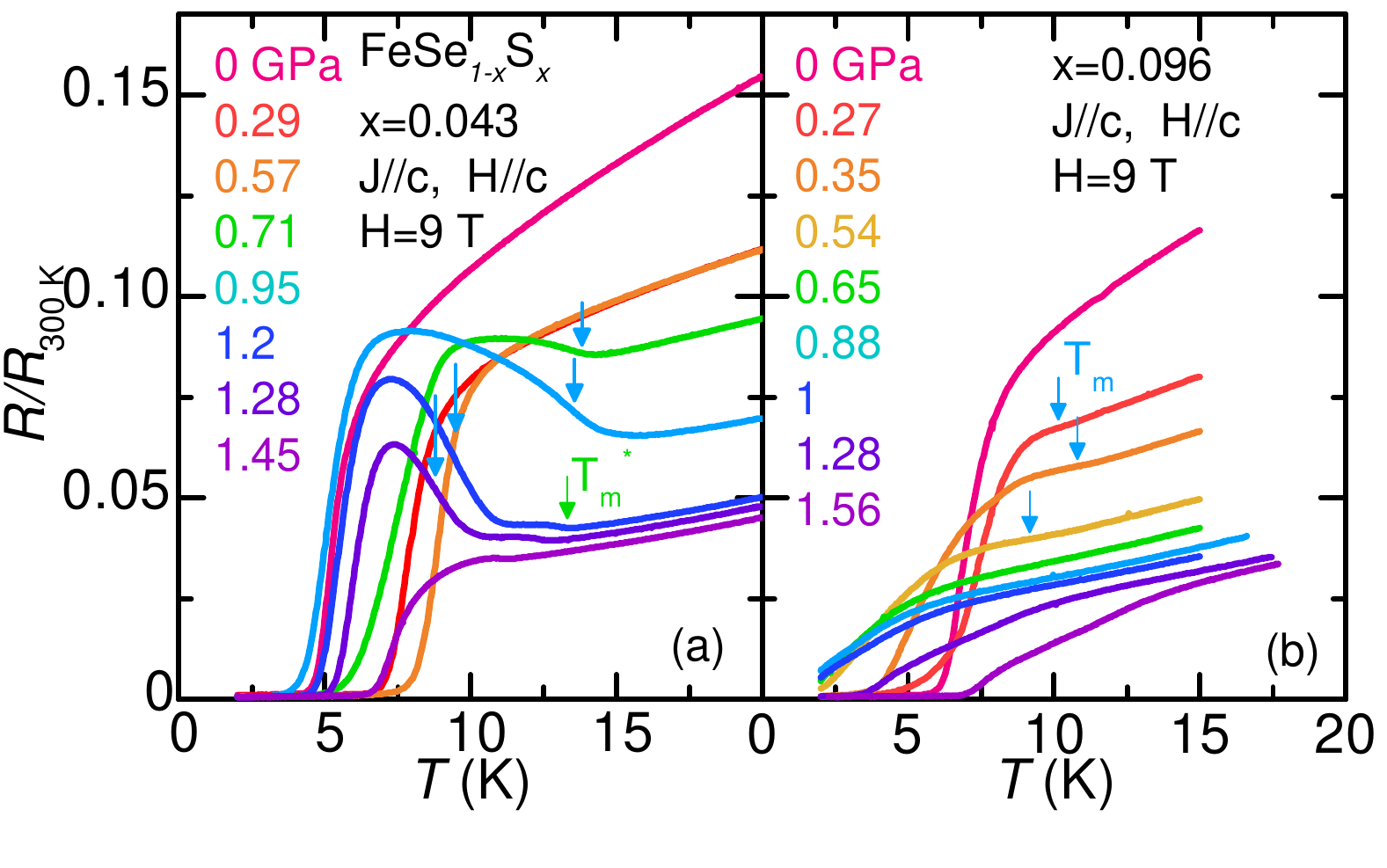}%
	\caption{Evolution of the temperature dependence of normalized resistance under various pressures with $H = 9$ T magnetic field applied parallel to $c$  axis for (a) $x=0.043$ and (b) $x=0.096$. For the $x=0.043$ sample, the magnetic phase transition indicated by blue arrows is more pronounced in field and a second anomaly is observed at slightly higher temperatures. This anomaly, at $T_m^*$, is indicated by green arrow and will be discussed in the Appendix. For $x=0.096$, magnetic field reveals a subtle anomaly between $0.27-0.54$ GPa, associated with the magnetic transition at $T_m$. 
		\label{raw_9T}}
\end{figure}
 
Figures 2-5 (a) show the pressure dependence of the resistance of Fe(Se$_{1-x}$S$_x$) for $x=0$, $x=0.043$, $0.096$ and $0.12$, respectively. In these plots the resistance is normalized by dividing it by the ambient-pressure, room-temperature value for each sample. 
In general, the resistance decreases under applied pressure. A non-monotonic change of the high-temperature resistance value for the $x=0.12$ sample is possibly due to contacting between the outside wiring and the piston cylinder pressure cell body in the first three pressure runs. The kink-like anomaly, associated with the structural phase transition $T_s$, is clearly visible in the lower pressure data and appears as a step-like anomaly in the temperature derivative $dR$/$dT$ (Figs. 2-5(c)). With increasing pressure, $T_s$ is suppressed in all compounds. The blow up of the low temperature region, presented in Figs. 2-5(b), highlights non-monotonic changes of $T_c$ under increasing pressure. Furthermore, the superconducting transition broadens systematically under pressure, a tendency observed in the parent compound in the magnetically ordered phase. The increasing broadening of the superconducting transition under pressure could also be due to inherent inhomogeneity of pressure when larger loads are applied and the substituted samples may be increasingly sensitive to this inhomogeneity.

The magnified scale in Figs 2-5 (b) reveals the effect of S-substitution on $T_m$. For $x=0.043$, an increase of resistance upon cooling is observed below 15 K for pressures between $0.71-1.03$ GPa. This anomaly is reminiscent of the resistance increase at $T_m$ of the parent compound at low pressures, shown in Fig. \ref{raw4}(b). We therefore associate it with the magnetic transition temperature $T_m$. In contrast to the parent compound, however, $T_m$ is much less prominent in the S-substituted samples.

A magnetic field suppresses $T_c$ but does not measurably affect $T_m$\cite{Kaluarachchi2016}, allowing for the study the magnetic transition in the absence of superconductivity. The application of a 9 T magnetic field, parallel to the $c$ axis, permits us to discern $T_m$ at pressures up to 1.28 GPa for the $x=0.043$ sample (Fig. \ref{raw_9T}). 
An additional anomaly at temperatures slightly above $T_m$ is observed for pressures greater than 0.95 GPa and is discussed in the appendix. 

No feature corresponding to a possible magnetic transition is observed in the resistance data for $x=0.096$ and $x=0.12$ in zero magnetic field. However, the application of a 9 T magnetic field reveals a subtle resistance anomaly between $0.27-0.54$ GPa for the $x=0.096$ sample (Fig. \ref{raw_9T}(b)), which may be associated with $T_m$. For the $x=0.12$ sample, even in a 9 T magnetic field, no anomaly that could be associated with magnetic ordering is observed in the resistance measurement with pressure up to 1.81 GPa. It is possible that the anomaly at $T_m$ is less pronounced in the in-plane resistance, which was measured for the $x=0.12$ sample, and therefore not resolved in these data.

%Temperature dependence of normalized resistance and derivative d\textit{R}/d\textit{T} taken at 0.95 GPa for x=0.043 compound. Criteria that is used to determine the structure transition $T_s$ and magnetic transition $T_m$ are shown by the dashed line.

\begin{figure}
	\includegraphics[width=8.6cm]{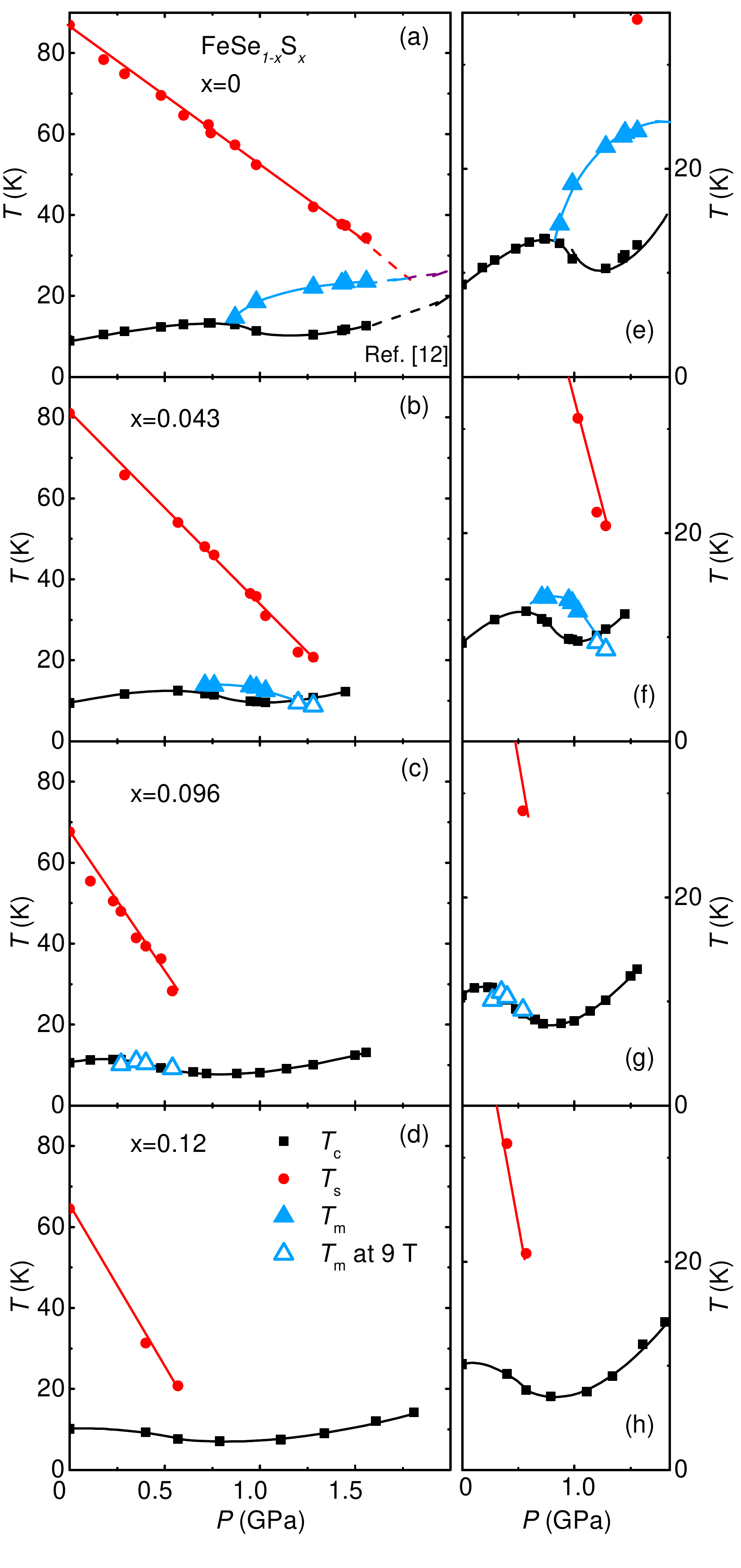}%
	\caption{Temperature - pressure phase diagrams of Fe(Se$_{1-x}$S$_{x}$) as determined from resistance measurements. The squares, circles and triangles circles represent the superconducting $T_c$, structural $T_s$ and magnetic $T_m$ phase transitions respectively. The solid lines are guides to the eye. Data in (a),(e) are taken from Ref. \onlinecite{Kaluarachchi2016}. The dashed lines in (a) represent extrapolations based of Refs. \onlinecite{Terashima2015,Kothapalli2016}. As shown in the left panels (a)-(d), for all compounds $T_s$ is suppressed linearly with increasing pressure. Panels (e)-(h) show the data on an expanded temperature scale. $T_c$ shows similar non-monotonic dependence on pressure with a local maximum and minimum. The magnetic order appears strongly suppressed upon substitution. The intersection of the $T_s$ lines and $T_c$ lines is not unique and does not coincide universally with either the minimum or the maximum of $T_c$.
		\label{TP_phase_diagram}}
\end{figure}

The values of $T_c$, $T_m$ and $T_s$ were obtained using the criteria outlined in Ref. \onlinecite{Kaluarachchi2016} and shown in Figs. \ref{raw4} and \ref{raw1}. $T_c$ is defined as the intersection between highest slope of $R(T)$ and zero resistance. $T_s$ is defined as the midpoint of the step in d$R$/d$T$, i.e., the midpoint of the kink in $R(T)$, and $T_m$ is defined as the point of the highest slope of the resistance. %Data for the parent compound $x=0$ were obtained using the same criteria \cite{Kaluarachchi2016}.
The resulting $p-T$ phase diagrams of Fe(Se$_{1-x}$S$_x$), $x=0-0.12$, are presented in Fig\,\ref{TP_phase_diagram}. 

\begin{figure}
	\includegraphics[width=8.6cm]{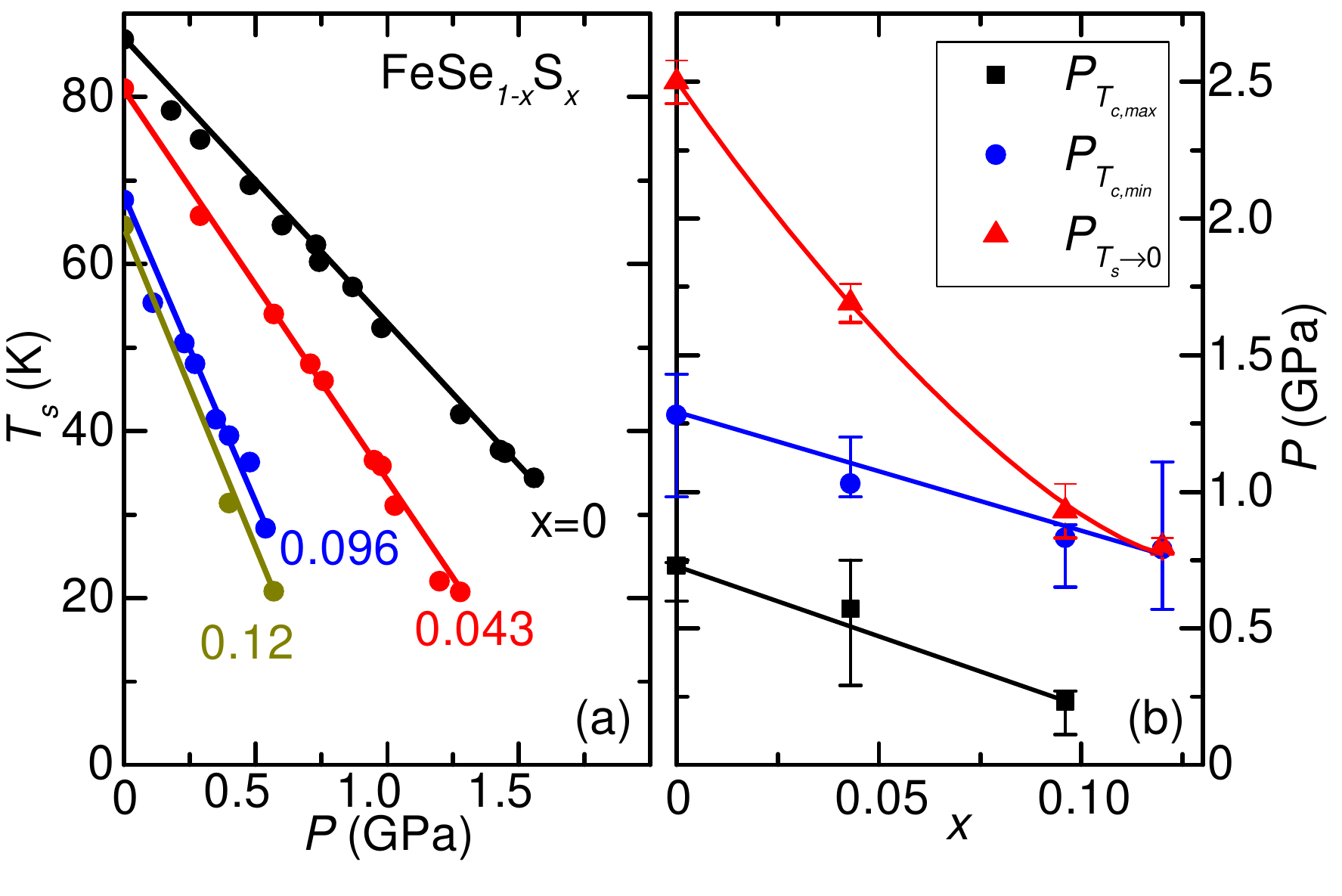}%
	\caption{(a) Pressure dependence of the structural transition temperature $T_s$ for Fe(Se$_{1-x}$S$_x$) with different substitution levels $x$. (b) Substitution dependence of the pressures  ${P_T}_{c,\mathrm{max}}$,  ${P_T}_{c,\mathrm{min}}$ and $P_{{T_s}\rightarrow0}$ which correspond to the local maximum of  $T_c$, minimum of $T_c$ and the extrapolation of $T_s$ to zero temperature, respectively. Solid lines are guides to the eyes. Data for $x=0$ are taken from Ref. \onlinecite{Kaluarachchi2016}.
		\label{Ts}}
\end{figure}

The orthorhombic phase line is clearly resolved in all of the phase diagrams in the pressure range below $\sim0.5-1.5$ GPa. At ambient pressure, $T_s$ is suppressed by 12\% S-substitution from 90 K to 60 K. Pressure suppresses $T_s$ almost linearly for all $x$, but as shown in Fig. \ref{Ts}(a), with increased rate $dT_s/dP$ for higher $x$.

For the parent compound FeSe, the magnetic transition at $T_m$ is observed for pressures greater than 0.8 GPa\cite{Kaluarachchi2016,Terashima2015}. Subsequent work has shown the magnetic phase to persist up to 6 GPa, with a dome-like dependence of $T_m$ on pressure\cite{Sun2016a,Sun2017}. For the $x=0.043$ sample, a similar phase line emerges above 0.5 GPa, and we tentatively associate it with $T_m$. But in contrast to pure FeSe, $T_m$ increases only slightly to a maximum of 13.8 K at 0.71 GPa and is suppressed to below $T_c$ already by 1.2 GPa. For higher S-content, $x=0.096$, this transition seems to occur within the small pressure range $0.27-0.57$ GPa and with a dome-like shape barely exceeding $T_c$ at its maximum. For $x=0.12$, no corresponding transition is resolved in the in-plane resistance measurement. 

For all measured substitution levels, $T_c$ of Fe(Se$_{1-x}$S$_x$) shows a similar non-monotonic dependence on pressure. The local maximum of $T_c$ shifts to lower pressure on increasing sulfur content, from ${P_T}_{c,\mathrm{max}}=0.73$ GPa for $x=0$ to 0.23 GPa for $x=0.096$ and close to ambient pressure for $x=0.12$. Likewise, the local minimum of $T_c$ shifts from ${P_T}_{c,\mathrm{min}}=1.28$ GPa for $x=0$ to 0.79 GPa for $x=0.12$, as presented in Fig. \ref{Ts}(b).

The clear suppression of $T_c$ below its local maximum in the intermediate pressure range is similar for all studied substitution levels. The onset of this suppression correlates with the emergence of the magnetic phase for $x=0-0.096$, even though in the $x=0.096$ sample, $T_m$ is indicated only by an extremely weak feature in resistivity and practically coincides with $T_c$. For $x=0.12$, $T_m$ is not visible at all. It seems likely that the competing order setting in at $T_m$ suppresses $T_c$ for $x=0-0.096$. However, whether this is still the case at higher substitution levels remains an open question and possibly another mechanism for the partial suppression of $T_c$ needs to be invoked. 

 The minimum of $T_c$ of pure FeSe at 1.3 GPa likely coincides with a change of the Fermi-surface under pressure\cite{Kaluarachchi2016,Terashima2016}. It is plausible that a similar change of Fermi surface occurs in the doped samples and is the origin of the local minimum of $T_c$. In contrast, the extrapolations of the $T_s$ phase lines intersect $T_c$ at non-unique positions for the different substitution levels. The extrapolation does not correlate universally with either the maximum or the minimum of $T_c$ in Fe(Se$_{1-x}$S$_x$), $x=0-0.12$ (Fig.\ref{Ts}(b)). This behavior differs from many other iron-based superconductor phase diagrams, where $T_s$ and $T_c$ typically intersect near the maximum of $T_c$ (Ref. \onlinecite{Paglione2010}). %In Fe(Se$_{1-x}$S$_x$) under pressure the extrapolation of the structural transition to zero does not correlate with either the maximum or the minimum of $T_c$. 

Fe(Se$_{0.904}$S$_{0.096}$) provides an example in which the structural transition extrapolates to the minimum of $T_c$. Several theories have discussed the influence of a nematic phase, and in particular of a nematic quantum critical point, on superconductivity\cite{Lederer2015,Labat2017}. In all cases, the nematic fluctuations are assumed to enhance (or induce) superconducting pairing and correlate with a maximum in $T_c$, opposite to the observed behavior. This is a sign that nematic fluctuations may not be involved in the superconducting pairing in this compound. %Note that even at ambient pressure, the extrapolation of $T_s\rightarrow0$ in Fe(Se$_{1-x}$S$_x$) does not coincide with a maximum of $T_c$ either.
% Another possibility is that an unidentified competing phase, or a change in Fermi-surface suppresses $T_c$, unrelated to the suppression of the nematic phase.  

The magnetic phase in the low-pressure range is extremely sensitive to S-substitution, but the orthorhombic/nematic phase is not. For example, in Fe(Se$_{0.957}$S$_{0.043}$) we observe only a tiny magnetic dome, contained entirely inside the nematic phase.  In pure FeSe, $T_m$ increases under applied pressure until $T_s$ and $T_m$ merge. The increase of orthorhombic distortion below $T_m$ in FeSe demonstrates the cooperative coupling of the two types of order\cite{Kothapalli2016}, similar to many iron-arsenide materials\cite{Kim2011}. In the well-known spin-nematic scenario for iron-arsenide materials\cite{Fernandes2012}, the nematic transition is believed to be a consequence of incipient stripe-type magnetic order. The strikingly different response of nematic and magnetic order to sulfur substitution in FeSe suggests, however, that the nematic phase in Fe(Se$_{1-x}$S$_x$) may not be related to the magnetic order observed in the low pressure range. A number of alternative scenarios for the origin of nematic order in FeSe have been put forward, including quadrupolar order\cite{Yu2015,Wang2016}, frustrated quantum paramagnetism\cite{Wang2015a} and a Pomeranchuk instability\cite{Chubukov2016}.

Isovalent substitution, as the replacement of selenium by sulfur, may be thought of as chemical pressure. Well-known examples in the iron-arsenide systems are BaFe$_2$(As$_{1-x}$P$_x$)$_2$ and Ba(Fe$_{1-x}$Ru$_x$)$_2$As$_2$\cite{Klintberg2010,Jiang2009,Colombier2009,Thaler2010}.
If pressure and substitution were simply additive, the $p-T$ phase diagrams for different substitution levels would be shifted with respect to each other. This is clearly not the case for the transition at $T_m$ in Fe(Se$_{1-x}$S$_x$), whose maximum temperature is strongly suppressed with increasing $x$. 
Sulfur substitution and pressure are not additive concerning $T_s$ either. Fig. \ref{Ts}(a) shows the $T_s$ phase lines for the four substitution levels $x=0$, $0.043$, $0.096$ and $x=0.12$. Both substitution and pressure suppress $T_s$, but the rate of suppression of $T_s$ under pressure depends on the substitution level. This would not be the case if S-substitution was simply additive to pressure. Similarly, an overlap of the "S-shaped" pressure dependence of $T_c$ for different $x$ can not be achieved by a simple shift. Even though ${P_T}_{c,\mathrm{max}}$ and ${P_T}_{c,\mathrm{min}}$ are suppressed at a similar rate by sulfur substitution (Fig.\ref{Ts}(b)), this "S" changes shape for increasing sulfur content. These comparisons demonstrate that sulfur substitution and physical pressure are not equivalent in FeSe concerning any phase transition and likely modify the electronic structure as well as any salient coupling constants in different ways.

 % Discussion of Tc, magnetic order and nematic quantum critical point. But interplay with magnetism may render it all more complicated. \cite{Lederer2015,Labat2017}. Magnetism seems to be correlated with decrease of Tc, but not increase

\section{Pressure-dependence of the upper critical field}

\begin{figure}
	\includegraphics[width=8.6cm]{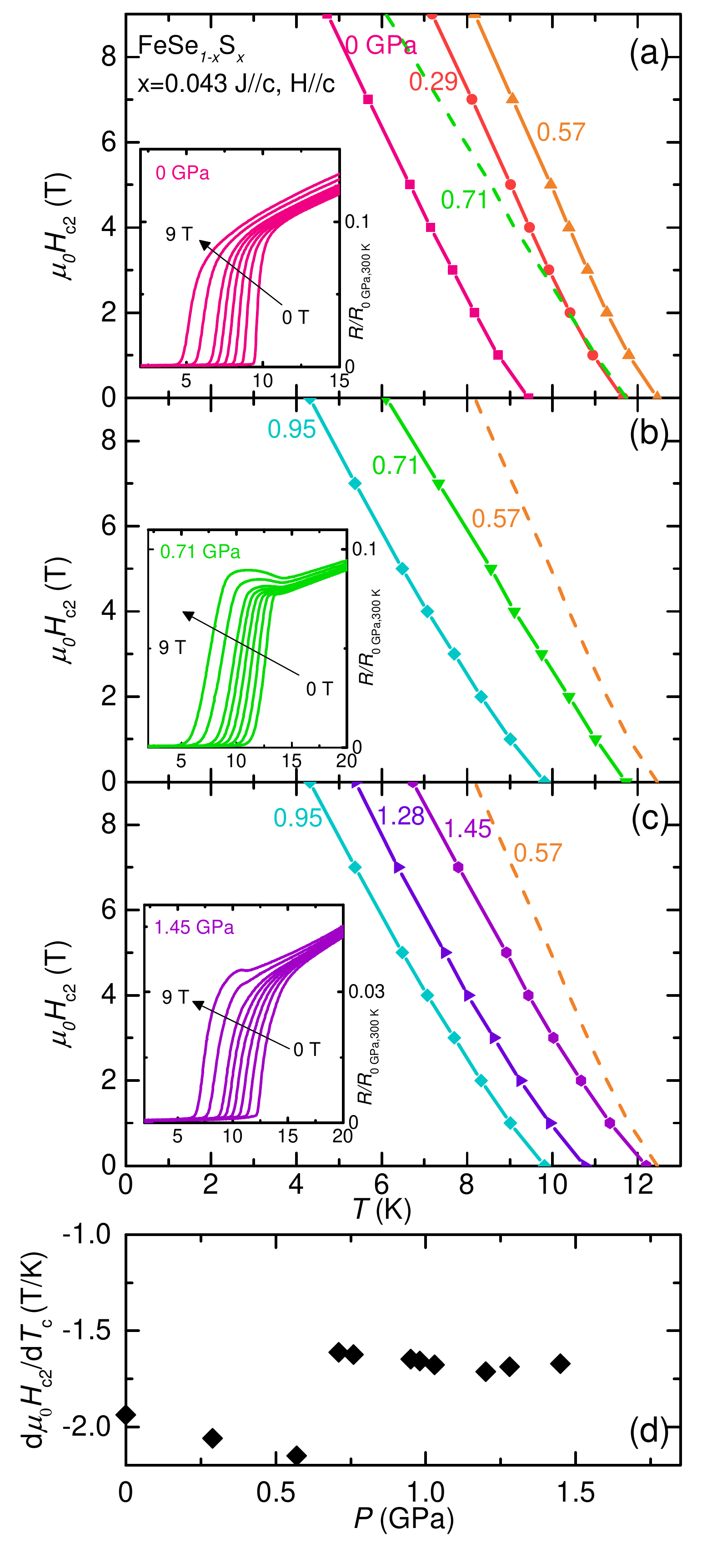}%
	\caption{Temperature dependence of the upper critical field $H_{c2}(\textit{T})$ measured in $j \parallel H \parallel c$ configuration under various pressures for the $x=0.043$ sample. Three regions are identified and separated by the local maximum and minimum of $T_c$ under pressure (panels (a), (b) and (c), respectively). A clear change of the $H_{c2}(T)$ slope is  observed between the first and second region only (panel (d)). Insets show representative resistance data under magnetic fields up to 9 T.
		\label{Hc2_Tc_print1}}
\end{figure}
\begin{figure}
	\includegraphics[width=8.6cm]{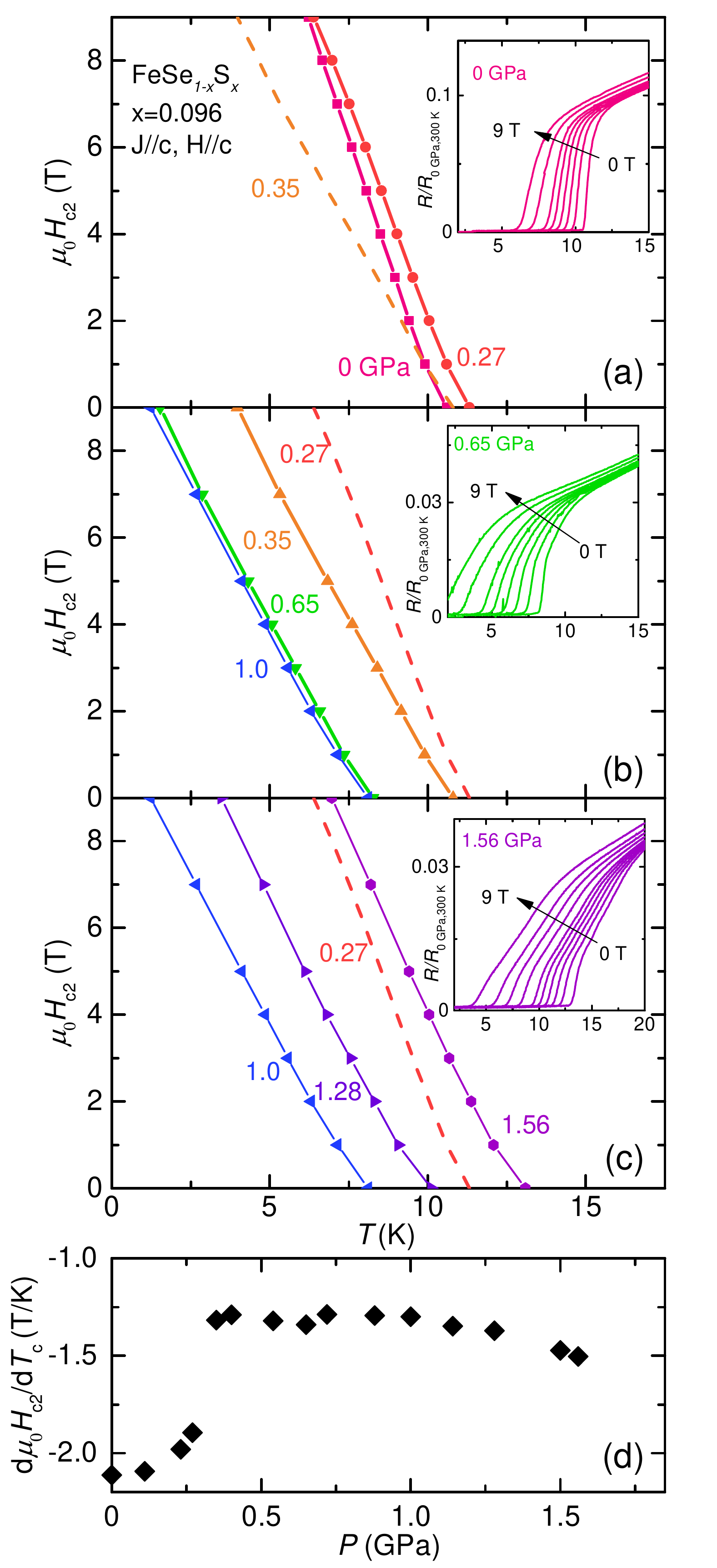}%
	\caption{Temperature dependence of the upper critical field $H_{c2}(\textit{T})$ measured in $j \parallel H \parallel c$configuration under various pressures for the $x=0.096$ sample. Three regions are identified and separated by the local maximum and minimum of $T_c$ under pressure (panels (a), (b) and (c), respectively). A clear change of the $H_{c2}(T)$ slope is  observed between the first and second region only (panel (d)). Insets show representative resistance data under magnetic fields up to 9 T.
		\label{Hc2_Tc_print2}}
\end{figure}
\begin{figure}
	\includegraphics[width=8.6cm]{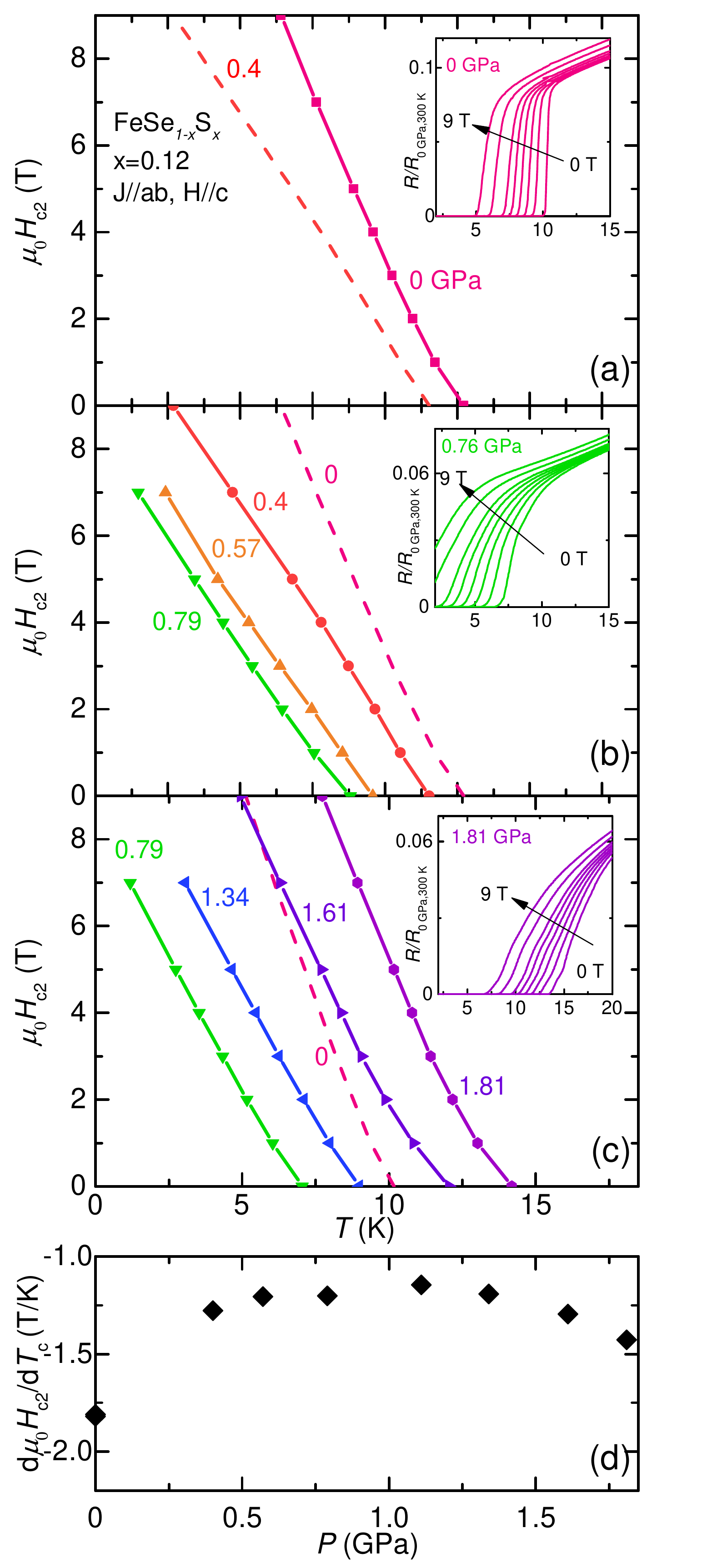}%
	\caption{Temperature dependence of the upper critical field $H_{c2}(\textit{T})$ measured in $H \parallel c$, $j \parallel ab$ configuration under various pressures for the $x=0.12$ sample. Three regions are identified and separated by the local maximum and minimum of $T_c$ under pressure (panels (a), (b) and (c), respectively). A clear change of the $H_{c2}(T)$ slope is  observed between the first and second region only (panel (d)). Insets show representative resistance data under magnetic fields up to 9 T.
		\label{Hc2_Tc_print3}}
\end{figure}

To better understand the superconducting properties of Fe(Se$_{1-x}$S$_x$), including the non-monotonic pressure dependence of $T_c$, the superconducting upper critical field is analyzed following Refs. \onlinecite{Kaluarachchi2016,Taufour2014}. Figs\,\ref{Hc2_Tc_print1}, \ref{Hc2_Tc_print2} and \ref{Hc2_Tc_print3} show the temperature dependence of the upper critical field $H_{c2}$ for $H||c$ of Fe(Se$_{1-x}$S$_x$) for $x=0.043$, $x=0.096$ and $x=0.12$ at various pressures. The insets show the temperature dependence of resistance in magnetic fields $H \parallel c$ between $0-9$ T, from which these data are obtained, for representative pressure values. Notably, for the $x=0.12$ sample, the current was applied along the ab plane, whereas the current was along the $c$-axis for the other compounds. In principle, the $j||H||c$ configuration can minimize the contribution of flux flow to the superconducting transitions, but no fundamental difference with different current directions  was observed between the measurements. At ambient pressure, the superconducting transition remains sharp for all field values. As the pressure is increased the superconducting transition becomes broader, especially in the $x=0.096$ and $x=0.12$ samples.

\begin{figure}
	\includegraphics[width=8.6cm]{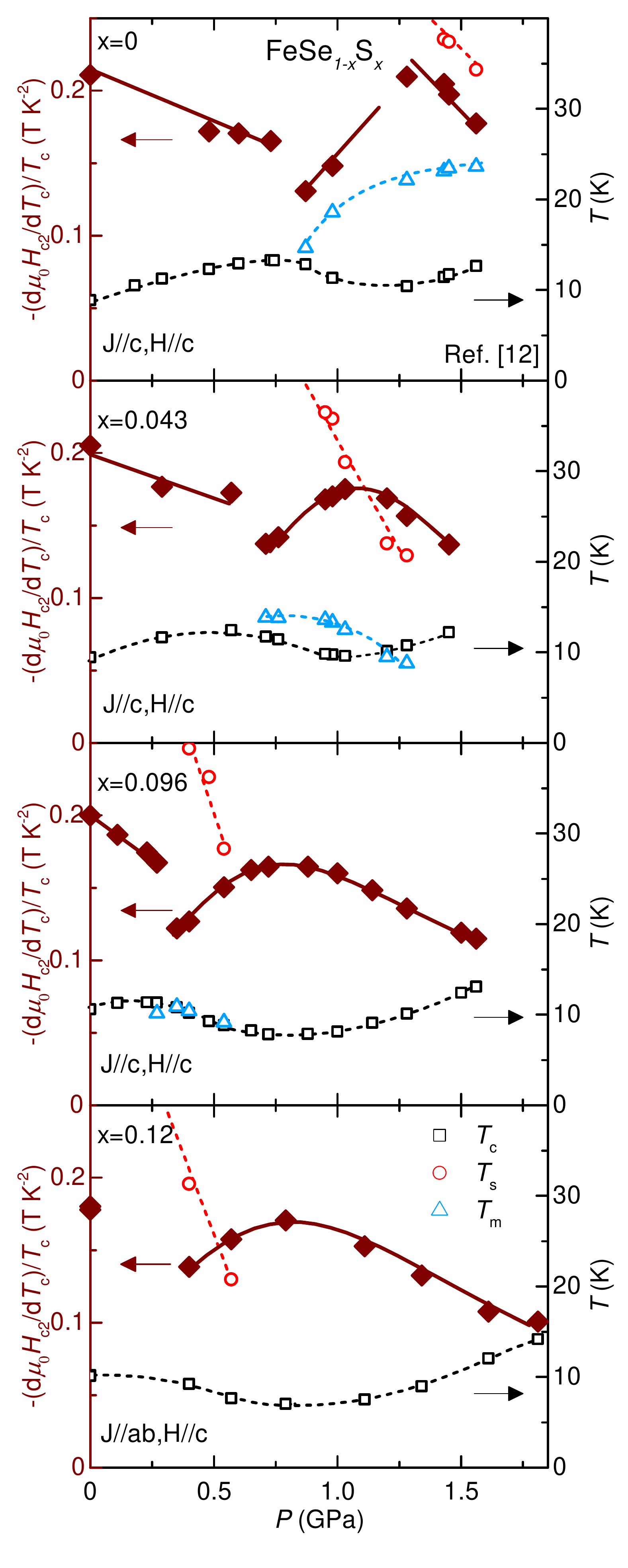}%
	\caption{Pressure dependence of the normalized upper critical field slope -[$d\mu_oH_{c2}$/$dT_c$]/$T_c$, plotted together with $T_c$, $T_s$ and $T_m$. For all compounds, an abrupt change of slope is observed near the local maximum of $T_c$. For the sulfur-containing compounds, a more continuous change of the slope occurs near the local minimum of $T_c$. Date in (a) is taken from Ref. \onlinecite{Kaluarachchi2016}.
		\label{Hc2_2_print}}
\end{figure}

A distinct change of the slope of $H_{c2}(T)$, which is abtained by fitting the 0-9 T date, is observed between 0.57 GPa and 0.71 GPa (between 0.27 GPa and 0.35 GPa) for $x=0.043$ ($x=0.096$). For $x=0.12$, a slope change occurs between ambient pressure and 0.4 GPa (Figures 9-11 (d)). These pressure ranges are close to the local maximum of $T_c$ and, for $x=0.043$ and $x=0.096$, the onset of magnetic order. No abrupt slope change of $H_{c2}$ occurs around the pressure associated local minimum of $T_c$.

 Fig\,\ref{Hc2_2_print} shows the pressure evolution of the upper critical field slope normalized by $T_c$, -[$d\mu_oH_{c2}$/$dT_c$]/$T_c$, and of the transition temperatures $T_c$, $T_s$, $T_m$ for $x=0$ (Ref. \onlinecite{Kaluarachchi2016}), $x=0.043$, $x=0.096$ and $x=0.12$. For all substitution levels, -[$d\mu_oH_{c2}$/$dT_c$]/$T_c$ exhibits a sudden decrease near the local maximum of $T_c$ under pressure. For the substituted compounds, a more continuous change is observed near the local minimum of $T_c$ at which point -[$d\mu_oH_{c2}$/$dT_c$]/$T_c$ has a broad maximum. 
 
Generally speaking, the slope of the upper critical field normalized by $T_c$, is related to the Fermi velocity and superconducting gap of the system\cite{Kogan2012}. In the clean limit for a single-band case,
\begin{equation}
 -[d\mu_oH_{c2}/dT_c]/T_c \propto 1/v_F^2,
 \label{eq:Hc2}
\end{equation} 
where $v_F$ is the Fermi velocity. Note that the mass enhancement expected at a quantum critical point should result in an increase of -[$d\mu_oH_{c2}$/$dT_c$]/$T_c$ (Ref. \onlinecite{Putzke2014}). The superconducting gap structure and, in a multiband-case, the coupling constants for the different bands are also involved\cite{Kogan2012}. A change of the normalized slope of $H_{c2}$ may result from changes of the Fermi surface, of the superconducting gap structure or of the pairing mechanism\cite{Taufour2014,Kogan2012}.  In addition, a change of scattering rates can also change $H_{c2}$ (Ref. \onlinecite{Kogan2014}). It was previously shown in pure FeSe that both the decrease of -[$d\mu_oH_{c2}$/$dT_c$]/$T_c$ close to the local maximum of $T_c$ as well as its increase close to the local minimum of $T_c$ under pressure can be explained by changes in the Fermi velocity \cite{Kaluarachchi2016}.

Similarly to pure FeSe, -[$d\mu_oH_{c2}$/$dT_c$]/$T_c$ of Fe(Se$_{1-x}$S$_x$) displays an abrupt decrease close to the local maximum of $T_c$ under pressure for all studied substitution levels. This points to a similar change of Fermi velocity as in the parent compound and supports the identification of this pressure level with the emergence of magnetic order entailing a reconstruction of the Fermi surface. Possibly, a change of electronic scattering rates at the onset of magnetic order also influences $H_{c2}$. The subsequent broad maximum of -$[d\mu_oH_{c2}/d T_c]/T_c$ results from dividing an almost pressure independent $d\mu_oH_{c2}/d T_c$ (Figs 9-11 (d)) by $T_c$, since $T_c$ displays a minimum in this pressure range. This maximum of the normalized slope of $H_{c2}$ may also be associated with a pressure-induced Fermi surface change or with a gradual mass enhancement at this pressure. Note that a pressure-independent 
$d\mu_oH_{c2}/d T_c$ indicates that $T_c \propto v_F^2$, according to equation \ref{eq:Hc2}.

\section{Conclusion}

\begin{figure}
	\includegraphics[width=8.6cm]{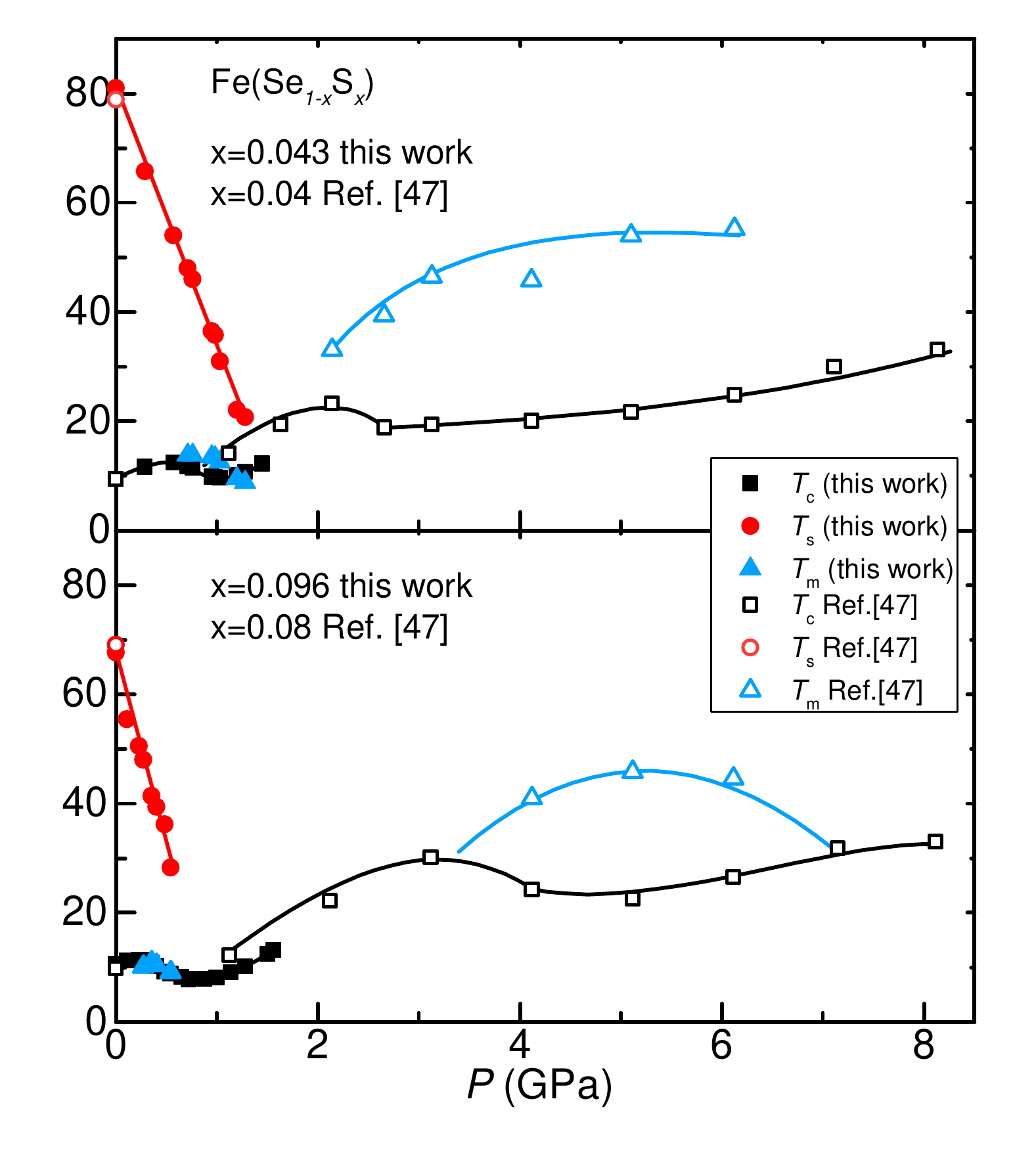}%
	\caption{Temperature - pressure phase diagrams of Fe(Se$_{1-x}$S$_{x}$) up to 8 GPa, including data from Ref. \onlinecite{Matsuura2017}. The squares, circles and triangles circles represent the superconducting $T_c$, structural $T_s$ and magnetic $T_m$ phase transitions respectively. Data from samples with similar $T_s$ value at ambient pressure are combined. The solid lines are guides to the eye. This represents an extension of our detailed low-pressure phase diagrams presented in Fig. \ref{TP_phase_diagram}. 
		\label{phase_diagram_combined}}.
\end{figure}

In conclusion, the resistance of sulfur-substituted FeSe$_{1-x}$S$_{x}$ ($x=0.043, 0.096, 0.12$) has been studied under pressures up to 1.8 GPa and in magnetic fields up to 9 T. $T_c$ exhibits a similar, non-monotonic pressure dependence with a local maximum and a local minimum for all substitution levels. $T_s$ is suppressed by pressure, at increasing rates for higher sulfur contents. The magnetic phase in the low-pressure range is strongly suppressed by substitution, which raises the question of how closely magnetic order and orthorhombic phase are related. Abrupt changes in the normalized slope of the upper critical field -[$d\mu_oH_{c2}$/$dT_c$]/$T_c$ near the local maximum of $T_c$ may indicate a Fermi-surface reconstruction coinciding with the emergence of magnetic order for $x=0-0.096$ and suggest its existence in $x=0.12$ as well. Another change of Fermi surface likely occurs near the local minimum of $T_c$ at slightly higher pressures. These results highlight the differences between chemical pressure and physical pressure as tuning parameters for FeSe.  

Note added: During the finalization of this manuscript, related results on the pressure-temperature phase diagrams of Fe(Se$_{1-x}$S$_{x}$) ($x=0.04-0.17$) with a focus on the higher pressure range 2-8 GPa were made available\cite{Matsuura2017}. By means of resistivity measurements in a cubic anvil cell, a prominent dome of likely magnetic order was found to exist in the higher pressure range, detached from the nematic phase for $x\geq0.04$. Taken together with the results presented here, this indicates that the pressure-temperature phase diagram of lightly S-substituted Fe(Se$_{1-x}$S$_x$) features two magnetic phases (see Fig. \ref{phase_diagram_combined}), possibly resulting from a splitting of the single pressure-induced magnetic dome of pure FeSe. The mechanism by such a splitting would occur remains to be studied, as indeed, the microscopic nature of the pressure-induced phases and their relation to each other. Altogether, the recent results reveal the astounding complexity of pressure- and substitution-tuned FeSe.

\begin{acknowledgements}
We would like to thank A.Kreyssig for useful discussions. This work was carried out at the Iowa State University and supported by the
Ames Laboratory, U.S. DOE, under Contract No. DE-AC02-
07CH11358. V.T. was partially supported by Critical Material
Institute, an Energy Innovation Hub funded by U.S. DOE,
Office of Energy Efficiency and Renewal Energy, Advanced
Manufacturing Office. L.X. was supported, in part by the W.M. Keck Foundation.
\end{acknowledgements}
\clearpage

\section{APPENDIX}
\begin{figure}
	\includegraphics[width=8.6cm]{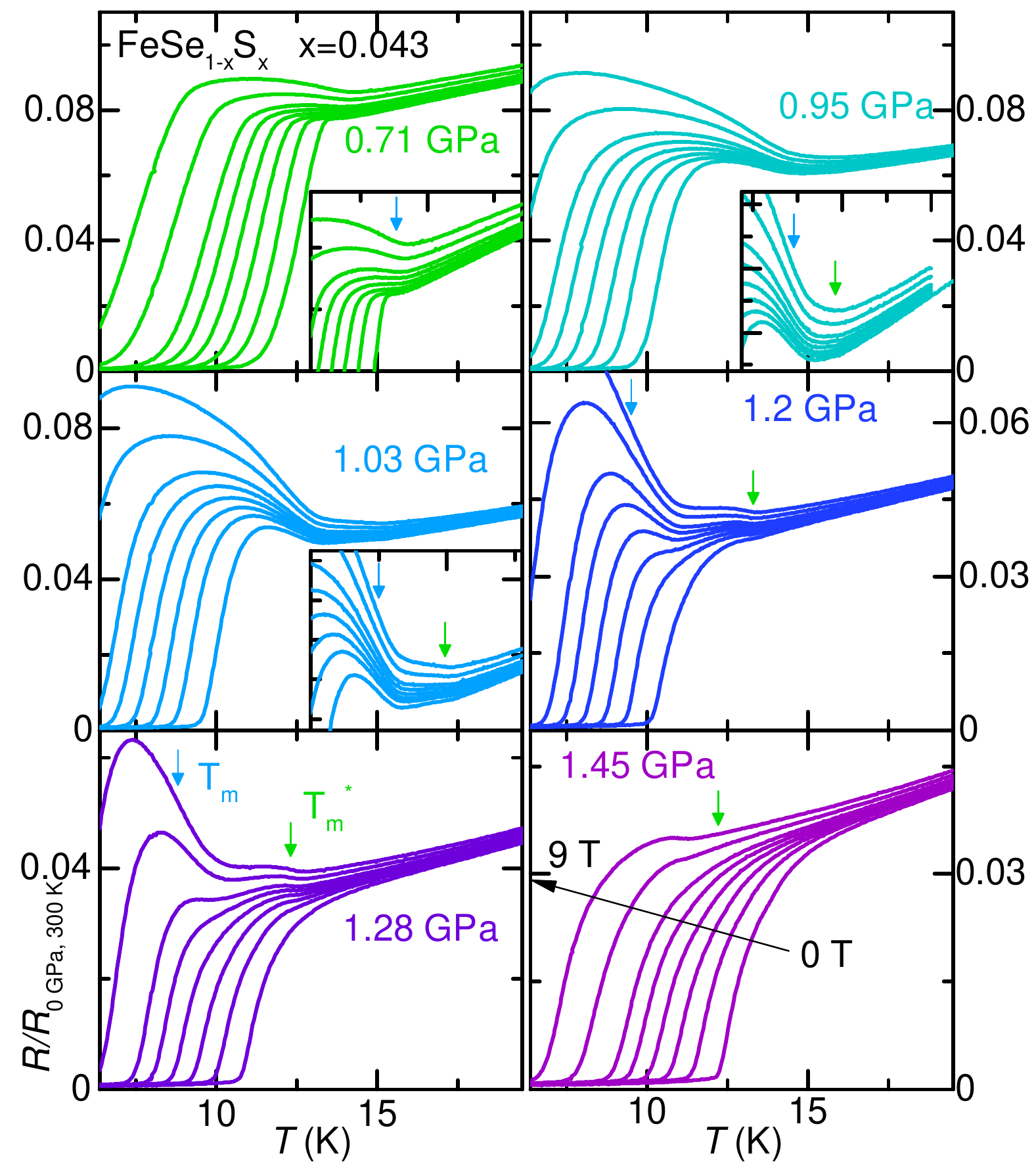}%
	\caption{Temperature dependence of the normalized resistance under magnetic field up to 9 T for selected pressures for compound FeSe$_{1-x}$S$_{x}$, $x=0.043$. Two anomalies associated with magnetic transition $T_m$ and possibly another magnetic transition $T_m^*$ are indicated by arrows.
		\label{Tswap_0p043}}
\end{figure}

An additional anomaly is observed in the resistance measurement for FeSe$_{0.957}$S$_{0.043}$ under pressure. As shown in Fig. \ref{Tswap_0p043}, in pressure range 0.95 - 1.45 GPa, two anomalies emerge above the superconducting transition. We associated the lower-temperature anomaly with the magnetic transition $T_m$ due to its similarities with the parent compound FeSe\cite{Sun2016a}. The other anomaly, labeled $T_m^*$, occurs slightly above $T_m$ and is indicated in Figs. \ref{raw_9T} and \ref{Tswap_0p043}. For 0.71 GPa, only $T_m$ is observed. From 0.95 - 1.03 GPa, both of these anomalies can be seen in zero field resistance measurements. Furthermore, with application of magnetic fields up to 9 T, these two anomalies barely shift. At higher pressures 1.2 - 1.45 GPa, those anomalies are no longer discernible in the zero field resistance measurements. However, by suppressing the superconducting transition with magnetic field, they are revealed in resistance for 1.2 GPa and 1.28 GPa. At our highest pressure of 1.45 GPa, only $T_m^*$ could be observed.

\begin{figure}
	\includegraphics[width=8.6cm]{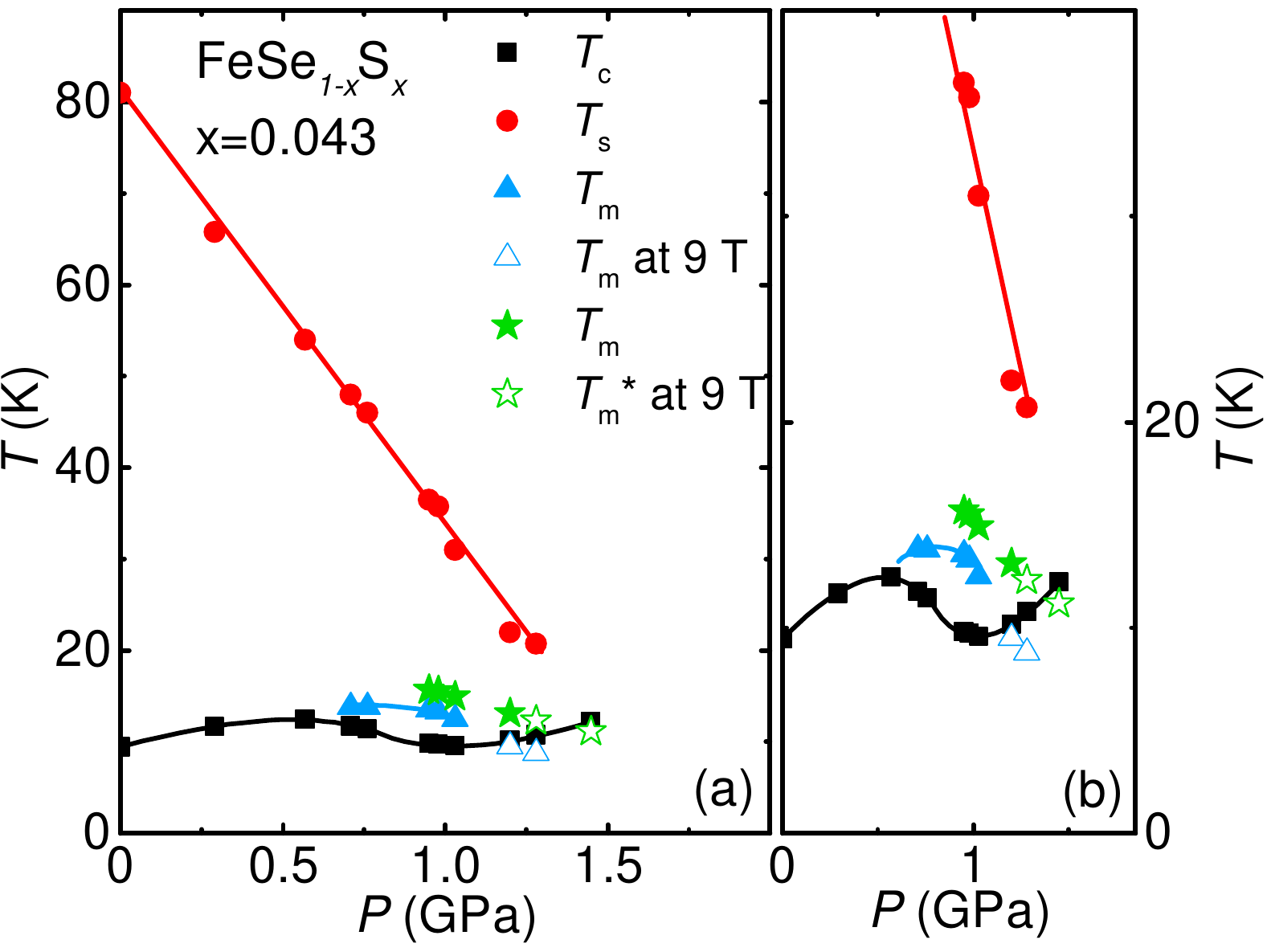}%
	\caption{Extended temperature - pressure phase diagram of FeSe$_{0.957}$S$_{0.043}$ as determined from resistance measurement as in Fig. \ref{TP_phase_diagram}. The $T_m^*$, indicated by green stars, represents the new anomaly we observed in this compound. The solid lines are guides for the eye.
		\label{TP_phase_diagram_9T_2}}
\end{figure}

The temperature - pressure phase diagram of FeSe$_{0.957}$S$_{0.043}$ complemented by including $T_m^*$ is presented in Fig. \ref{TP_phase_diagram_9T_2}. $T_m$ exhibits a dome-like pressure dependence, whereas $T_m^*$ emerges on the high-pressure side of this dome. Whether this new anomaly $T_m^*$ is related to a possible incommensurate magnetic transition or a different phase transition needs further studies.

%\begin{figure}
%	\includegraphics[width=8.6cm]{Tswap_0p043}%
%	\caption{
%		\label{RT_raw_9T}}
%\end{figure}

\clearpage
%\begin{figure}
%	\includegraphics[width=8.6cm]{Hc2_Tc_print3}%
%	\caption{Temperature dependence of the upper critical field $H_{c2,c}(\textit{T})$ measured in $H \parallel c$ configuration under various pressures for x=0.12 compound.
%			\label{Hc2_Tc_print3}}
%\end{figure}

%\begin{figure}
%	\includegraphics[width=8.6cm]{Hc2_2_print1}%
%	\caption{Pressure dependence of the normalized slope $-[d H_{c2,c}/d T_c]/T_c$ plotted together with $T_c$. For all compounds, an abrupt change of slope is observed near the local maximum of $T_c$. For doped compounds, the change of the slope is more continuous near the local minimum of $T_c$.
%		\label{Hc2_2_print1}}
%\end{figure}

%\bibliographystyle{apsrev4-1}
%\bibliography{references} 
%\bibliography{MyRef}

\begin{thebibliography}{47}%
	\makeatletter
	\providecommand \@ifxundefined [1]{%
		\@ifx{#1\undefined}
	}%
	\providecommand \@ifnum [1]{%
		\ifnum #1\expandafter \@firstoftwo
		\else \expandafter \@secondoftwo
		\fi
	}%
	\providecommand \@ifx [1]{%
		\ifx #1\expandafter \@firstoftwo
		\else \expandafter \@secondoftwo
		\fi
	}%
	\providecommand \natexlab [1]{#1}%
	\providecommand \enquote  [1]{``#1''}%
	\providecommand \bibnamefont  [1]{#1}%
	\providecommand \bibfnamefont [1]{#1}%
	\providecommand \citenamefont [1]{#1}%
	\providecommand \href@noop [0]{\@secondoftwo}%
	\providecommand \href [0]{\begingroup \@sanitize@url \@href}%
	\providecommand \@href[1]{\@@startlink{#1}\@@href}%
	\providecommand \@@href[1]{\endgroup#1\@@endlink}%
	\providecommand \@sanitize@url [0]{\catcode `\\12\catcode `\$12\catcode
		`\&12\catcode `\#12\catcode `\^12\catcode `\_12\catcode `\%12\relax}%
	\providecommand \@@startlink[1]{}%
	\providecommand \@@endlink[0]{}%
	\providecommand \url  [0]{\begingroup\@sanitize@url \@url }%
	\providecommand \@url [1]{\endgroup\@href {#1}{\urlprefix }}%
	\providecommand \urlprefix  [0]{URL }%
	\providecommand \Eprint [0]{\href }%
	\providecommand \doibase [0]{http://dx.doi.org/}%
	\providecommand \selectlanguage [0]{\@gobble}%
	\providecommand \bibinfo  [0]{\@secondoftwo}%
	\providecommand \bibfield  [0]{\@secondoftwo}%
	\providecommand \translation [1]{[#1]}%
	\providecommand \BibitemOpen [0]{}%
	\providecommand \bibitemStop [0]{}%
	\providecommand \bibitemNoStop [0]{.\EOS\space}%
	\providecommand \EOS [0]{\spacefactor3000\relax}%
	\providecommand \BibitemShut  [1]{\csname bibitem#1\endcsname}%
	\let\auto@bib@innerbib\@empty
	%</preamble>
	\bibitem [{\citenamefont {Canfield}\ and\ \citenamefont
		{Bud'ko}(2010)}]{Canfield2010}%
	\BibitemOpen
	\bibfield  {author} {\bibinfo {author} {\bibfnamefont {P.~C.}\ \bibnamefont
			{Canfield}}\ and\ \bibinfo {author} {\bibfnamefont {S.~L.}\ \bibnamefont
			{Bud'ko}},\ }\href {\doibase 10.1146/annurev-conmatphys-070909-104041}
	{\bibfield  {journal} {\bibinfo  {journal} {Annu. Rev. Condens. Matter
				Phys.}\ }\textbf {\bibinfo {volume} {1}},\ \bibinfo {pages} {27} (\bibinfo
		{year} {2010})}\BibitemShut {NoStop}%
	\bibitem [{\citenamefont {Paglione}\ and\ \citenamefont
		{Greene}(2010)}]{Paglione2010}%
	\BibitemOpen
	\bibfield  {author} {\bibinfo {author} {\bibfnamefont {J.}~\bibnamefont
			{Paglione}}\ and\ \bibinfo {author} {\bibfnamefont {R.~L.}\ \bibnamefont
			{Greene}},\ }\href {http://dx.doi.org/10.1038/nphys1759} {\bibfield
		{journal} {\bibinfo  {journal} {Nat Phys}\ }\textbf {\bibinfo {volume} {6}},\
		\bibinfo {pages} {645} (\bibinfo {year} {2010})}\BibitemShut {NoStop}%
	\bibitem [{\citenamefont {Fernandes}\ \emph {et~al.}(2014)\citenamefont
		{Fernandes}, \citenamefont {Chubukov},\ and\ \citenamefont
		{Schmalian}}]{Fernandes2014}%
	\BibitemOpen
	\bibfield  {author} {\bibinfo {author} {\bibfnamefont {R.~M.}\ \bibnamefont
			{Fernandes}}, \bibinfo {author} {\bibfnamefont {A.~V.}\ \bibnamefont
			{Chubukov}}, \ and\ \bibinfo {author} {\bibfnamefont {J.}~\bibnamefont
			{Schmalian}},\ }\href {http://dx.doi.org/10.1038/nphys2877} {\bibfield
		{journal} {\bibinfo  {journal} {Nat Phys}\ }\textbf {\bibinfo {volume}
			{10}},\ \bibinfo {pages} {97} (\bibinfo {year} {2014})}\BibitemShut {NoStop}%
	\bibitem [{\citenamefont {Kasahara}\ \emph {et~al.}(2010)\citenamefont
		{Kasahara}, \citenamefont {Shibauchi}, \citenamefont {Hashimoto},
		\citenamefont {Ikada}, \citenamefont {Tonegawa}, \citenamefont {Okazaki},
		\citenamefont {Shishido}, \citenamefont {Ikeda}, \citenamefont {Takeya},
		\citenamefont {Hirata}, \citenamefont {Terashima},\ and\ \citenamefont
		{Matsuda}}]{Kasahara2010}%
	\BibitemOpen
	\bibfield  {author} {\bibinfo {author} {\bibfnamefont {S.}~\bibnamefont
			{Kasahara}}, \bibinfo {author} {\bibfnamefont {T.}~\bibnamefont {Shibauchi}},
		\bibinfo {author} {\bibfnamefont {K.}~\bibnamefont {Hashimoto}}, \bibinfo
		{author} {\bibfnamefont {K.}~\bibnamefont {Ikada}}, \bibinfo {author}
		{\bibfnamefont {S.}~\bibnamefont {Tonegawa}}, \bibinfo {author}
		{\bibfnamefont {R.}~\bibnamefont {Okazaki}}, \bibinfo {author} {\bibfnamefont
			{H.}~\bibnamefont {Shishido}}, \bibinfo {author} {\bibfnamefont
			{H.}~\bibnamefont {Ikeda}}, \bibinfo {author} {\bibfnamefont
			{H.}~\bibnamefont {Takeya}}, \bibinfo {author} {\bibfnamefont
			{K.}~\bibnamefont {Hirata}}, \bibinfo {author} {\bibfnamefont
			{T.}~\bibnamefont {Terashima}}, \ and\ \bibinfo {author} {\bibfnamefont
			{Y.}~\bibnamefont {Matsuda}},\ }\href {\doibase 10.1103/PhysRevB.81.184519}
	{\bibfield  {journal} {\bibinfo  {journal} {Phys. Rev. B}\ }\textbf {\bibinfo
			{volume} {81}},\ \bibinfo {pages} {184519} (\bibinfo {year}
		{2010})}\BibitemShut {NoStop}%
	\bibitem [{\citenamefont {Putzke}\ \emph {et~al.}(2014)\citenamefont {Putzke},
		\citenamefont {Walmsley}, \citenamefont {Fletcher}, \citenamefont {Malone},
		\citenamefont {Vignolles}, \citenamefont {Proust}, \citenamefont {Badoux},
		\citenamefont {See}, \citenamefont {Beere}, \citenamefont {Ritchie},
		\citenamefont {Kasahara}, \citenamefont {Mizukami}, \citenamefont
		{Shibauchi}, \citenamefont {Matsuda},\ and\ \citenamefont
		{Carrington}}]{Putzke2014}%
	\BibitemOpen
	\bibfield  {author} {\bibinfo {author} {\bibfnamefont {C.}~\bibnamefont
			{Putzke}}, \bibinfo {author} {\bibfnamefont {P.}~\bibnamefont {Walmsley}},
		\bibinfo {author} {\bibfnamefont {J.~D.}\ \bibnamefont {Fletcher}}, \bibinfo
		{author} {\bibfnamefont {L.}~\bibnamefont {Malone}}, \bibinfo {author}
		{\bibfnamefont {D.}~\bibnamefont {Vignolles}}, \bibinfo {author}
		{\bibfnamefont {C.}~\bibnamefont {Proust}}, \bibinfo {author} {\bibfnamefont
			{S.}~\bibnamefont {Badoux}}, \bibinfo {author} {\bibfnamefont
			{P.}~\bibnamefont {See}}, \bibinfo {author} {\bibfnamefont {H.~E.}\
			\bibnamefont {Beere}}, \bibinfo {author} {\bibfnamefont {D.~A.}\ \bibnamefont
			{Ritchie}}, \bibinfo {author} {\bibfnamefont {S.}~\bibnamefont {Kasahara}},
		\bibinfo {author} {\bibfnamefont {Y.}~\bibnamefont {Mizukami}}, \bibinfo
		{author} {\bibfnamefont {T.}~\bibnamefont {Shibauchi}}, \bibinfo {author}
		{\bibfnamefont {Y.}~\bibnamefont {Matsuda}}, \ and\ \bibinfo {author}
		{\bibfnamefont {A.}~\bibnamefont {Carrington}},\ }\href
	{http://dx.doi.org/10.1038/ncomms6679} {\bibfield  {journal} {\bibinfo
			{journal} {Nat. Commun.}\ }\textbf {\bibinfo {volume} {5}},\ \bibinfo {pages}
		{5679} (\bibinfo {year} {2014})}\BibitemShut {NoStop}%
	\bibitem [{\citenamefont {McQueen}\ \emph {et~al.}(2009)\citenamefont
		{McQueen}, \citenamefont {Williams}, \citenamefont {Stephens}, \citenamefont
		{Tao}, \citenamefont {Zhu}, \citenamefont {Ksenofontov}, \citenamefont
		{Casper}, \citenamefont {Felser},\ and\ \citenamefont {Cava}}]{McQueen2009}%
	\BibitemOpen
	\bibfield  {author} {\bibinfo {author} {\bibfnamefont {T.~M.}\ \bibnamefont
			{McQueen}}, \bibinfo {author} {\bibfnamefont {A.~J.}\ \bibnamefont
			{Williams}}, \bibinfo {author} {\bibfnamefont {P.~W.}\ \bibnamefont
			{Stephens}}, \bibinfo {author} {\bibfnamefont {J.}~\bibnamefont {Tao}},
		\bibinfo {author} {\bibfnamefont {Y.}~\bibnamefont {Zhu}}, \bibinfo {author}
		{\bibfnamefont {V.}~\bibnamefont {Ksenofontov}}, \bibinfo {author}
		{\bibfnamefont {F.}~\bibnamefont {Casper}}, \bibinfo {author} {\bibfnamefont
			{C.}~\bibnamefont {Felser}}, \ and\ \bibinfo {author} {\bibfnamefont {R.~J.}\
			\bibnamefont {Cava}},\ }\href {\doibase 10.1103/PhysRevLett.103.057002}
	{\bibfield  {journal} {\bibinfo  {journal} {Phys. Rev. Lett.}\ }\textbf
		{\bibinfo {volume} {103}},\ \bibinfo {pages} {057002} (\bibinfo {year}
		{2009})}\BibitemShut {NoStop}%
	\bibitem [{\citenamefont {Bendele}\ \emph {et~al.}(2010)\citenamefont
		{Bendele}, \citenamefont {Amato}, \citenamefont {Conder}, \citenamefont
		{Elender}, \citenamefont {Keller}, \citenamefont {Klauss}, \citenamefont
		{Luetkens}, \citenamefont {Pomjakushina}, \citenamefont {Raselli},\ and\
		\citenamefont {Khasanov}}]{Bendele2010}%
	\BibitemOpen
	\bibfield  {author} {\bibinfo {author} {\bibfnamefont {M.}~\bibnamefont
			{Bendele}}, \bibinfo {author} {\bibfnamefont {A.}~\bibnamefont {Amato}},
		\bibinfo {author} {\bibfnamefont {K.}~\bibnamefont {Conder}}, \bibinfo
		{author} {\bibfnamefont {M.}~\bibnamefont {Elender}}, \bibinfo {author}
		{\bibfnamefont {H.}~\bibnamefont {Keller}}, \bibinfo {author} {\bibfnamefont
			{H.-H.}\ \bibnamefont {Klauss}}, \bibinfo {author} {\bibfnamefont
			{H.}~\bibnamefont {Luetkens}}, \bibinfo {author} {\bibfnamefont
			{E.}~\bibnamefont {Pomjakushina}}, \bibinfo {author} {\bibfnamefont
			{A.}~\bibnamefont {Raselli}}, \ and\ \bibinfo {author} {\bibfnamefont
			{R.}~\bibnamefont {Khasanov}},\ }\href {\doibase
		10.1103/PhysRevLett.104.087003} {\bibfield  {journal} {\bibinfo  {journal}
			{Phys. Rev. Lett.}\ }\textbf {\bibinfo {volume} {104}},\ \bibinfo {pages}
		{087003} (\bibinfo {year} {2010})}\BibitemShut {NoStop}%
	\bibitem [{\citenamefont {Chen}\ \emph {et~al.}(2017)\citenamefont {Chen},
		\citenamefont {Zhu}, \citenamefont {Yang},\ and\ \citenamefont
		{Wen}}]{Chen2017}%
	\BibitemOpen
	\bibfield  {author} {\bibinfo {author} {\bibfnamefont {G.}~\bibnamefont
			{Chen}}, \bibinfo {author} {\bibfnamefont {X.}~\bibnamefont {Zhu}}, \bibinfo
		{author} {\bibfnamefont {H.}~\bibnamefont {Yang}}, \ and\ \bibinfo {author}
		{\bibfnamefont {H.}~\bibnamefont {Wen}},\ }\href
	{https://arxiv.org/abs/1703.08680} {\bibfield  {journal} {\bibinfo  {journal}
			{arXiv:1703.08680}\ } (\bibinfo {year} {2017})}\BibitemShut {NoStop}%
	\bibitem [{\citenamefont {Lin}\ \emph {et~al.}(2011)\citenamefont {Lin},
		\citenamefont {Hsieh}, \citenamefont {Chareev}, \citenamefont {Vasiliev},
		\citenamefont {Parsons},\ and\ \citenamefont {Yang}}]{Lin2011}%
	\BibitemOpen
	\bibfield  {author} {\bibinfo {author} {\bibfnamefont {J.-Y.}\ \bibnamefont
			{Lin}}, \bibinfo {author} {\bibfnamefont {Y.~S.}\ \bibnamefont {Hsieh}},
		\bibinfo {author} {\bibfnamefont {D.~A.}\ \bibnamefont {Chareev}}, \bibinfo
		{author} {\bibfnamefont {A.~N.}\ \bibnamefont {Vasiliev}}, \bibinfo {author}
		{\bibfnamefont {Y.}~\bibnamefont {Parsons}}, \ and\ \bibinfo {author}
		{\bibfnamefont {H.~D.}\ \bibnamefont {Yang}},\ }\href {\doibase
		10.1103/PhysRevB.84.220507} {\bibfield  {journal} {\bibinfo  {journal} {Phys.
				Rev. B}\ }\textbf {\bibinfo {volume} {84}},\ \bibinfo {pages} {220507}
		(\bibinfo {year} {2011})}\BibitemShut {NoStop}%
	\bibitem [{\citenamefont {Bendele}\ \emph {et~al.}(2012)\citenamefont
		{Bendele}, \citenamefont {Ichsanow}, \citenamefont {Pashkevich},
		\citenamefont {Keller}, \citenamefont {Str\"assle}, \citenamefont {Gusev},
		\citenamefont {Pomjakushina}, \citenamefont {Conder}, \citenamefont
		{Khasanov},\ and\ \citenamefont {Keller}}]{Bendele2012}%
	\BibitemOpen
	\bibfield  {author} {\bibinfo {author} {\bibfnamefont {M.}~\bibnamefont
			{Bendele}}, \bibinfo {author} {\bibfnamefont {A.}~\bibnamefont {Ichsanow}},
		\bibinfo {author} {\bibfnamefont {Y.}~\bibnamefont {Pashkevich}}, \bibinfo
		{author} {\bibfnamefont {L.}~\bibnamefont {Keller}}, \bibinfo {author}
		{\bibfnamefont {T.}~\bibnamefont {Str\"assle}}, \bibinfo {author}
		{\bibfnamefont {A.}~\bibnamefont {Gusev}}, \bibinfo {author} {\bibfnamefont
			{E.}~\bibnamefont {Pomjakushina}}, \bibinfo {author} {\bibfnamefont
			{K.}~\bibnamefont {Conder}}, \bibinfo {author} {\bibfnamefont
			{R.}~\bibnamefont {Khasanov}}, \ and\ \bibinfo {author} {\bibfnamefont
			{H.}~\bibnamefont {Keller}},\ }\href {\doibase 10.1103/PhysRevB.85.064517}
	{\bibfield  {journal} {\bibinfo  {journal} {Phys. Rev. B}\ }\textbf {\bibinfo
			{volume} {85}},\ \bibinfo {pages} {064517} (\bibinfo {year}
		{2012})}\BibitemShut {NoStop}%
	\bibitem [{\citenamefont {Terashima}\ \emph {et~al.}(2015)\citenamefont
		{Terashima}, \citenamefont {Kikugawa}, \citenamefont {Kasahara},
		\citenamefont {Watashige}, \citenamefont {Shibauchi}, \citenamefont
		{Matsuda}, \citenamefont {Wolf}, \citenamefont {B\"ohmer}, \citenamefont
		{Hardy}, \citenamefont {Meingast}, \citenamefont {v.~L\"ohneysen},\ and\
		\citenamefont {Uji}}]{Terashima2015}%
	\BibitemOpen
	\bibfield  {author} {\bibinfo {author} {\bibfnamefont {T.}~\bibnamefont
			{Terashima}}, \bibinfo {author} {\bibfnamefont {N.}~\bibnamefont {Kikugawa}},
		\bibinfo {author} {\bibfnamefont {S.}~\bibnamefont {Kasahara}}, \bibinfo
		{author} {\bibfnamefont {T.}~\bibnamefont {Watashige}}, \bibinfo {author}
		{\bibfnamefont {T.}~\bibnamefont {Shibauchi}}, \bibinfo {author}
		{\bibfnamefont {Y.}~\bibnamefont {Matsuda}}, \bibinfo {author} {\bibfnamefont
			{T.}~\bibnamefont {Wolf}}, \bibinfo {author} {\bibfnamefont {A.~E.}\
			\bibnamefont {B\"ohmer}}, \bibinfo {author} {\bibfnamefont {F.}~\bibnamefont
			{Hardy}}, \bibinfo {author} {\bibfnamefont {C.}~\bibnamefont {Meingast}},
		\bibinfo {author} {\bibfnamefont {H.}~\bibnamefont {v.~L\"ohneysen}}, \ and\
		\bibinfo {author} {\bibfnamefont {S.}~\bibnamefont {Uji}},\ }\href {\doibase
		10.7566/JPSJ.84.063701} {\bibfield  {journal} {\bibinfo  {journal} {J. Phys.
				Soc. Jpn.}\ }\textbf {\bibinfo {volume} {84}},\ \bibinfo {pages} {063701}
		(\bibinfo {year} {2015})}\BibitemShut {NoStop}%
	\bibitem [{\citenamefont {Kaluarachchi}\ \emph {et~al.}(2016)\citenamefont
		{Kaluarachchi}, \citenamefont {Taufour}, \citenamefont {B\"ohmer},
		\citenamefont {Tanatar}, \citenamefont {Bud'ko}, \citenamefont {Kogan},
		\citenamefont {Prozorov},\ and\ \citenamefont {Canfield}}]{Kaluarachchi2016}%
	\BibitemOpen
	\bibfield  {author} {\bibinfo {author} {\bibfnamefont {U.~S.}\ \bibnamefont
			{Kaluarachchi}}, \bibinfo {author} {\bibfnamefont {V.}~\bibnamefont
			{Taufour}}, \bibinfo {author} {\bibfnamefont {A.~E.}\ \bibnamefont
			{B\"ohmer}}, \bibinfo {author} {\bibfnamefont {M.~A.}\ \bibnamefont
			{Tanatar}}, \bibinfo {author} {\bibfnamefont {S.~L.}\ \bibnamefont {Bud'ko}},
		\bibinfo {author} {\bibfnamefont {V.~G.}\ \bibnamefont {Kogan}}, \bibinfo
		{author} {\bibfnamefont {R.}~\bibnamefont {Prozorov}}, \ and\ \bibinfo
		{author} {\bibfnamefont {P.~C.}\ \bibnamefont {Canfield}},\ }\href {\doibase
		10.1103/PhysRevB.93.064503} {\bibfield  {journal} {\bibinfo  {journal} {Phys.
				Rev. B}\ }\textbf {\bibinfo {volume} {93}},\ \bibinfo {pages} {064503}
		(\bibinfo {year} {2016})}\BibitemShut {NoStop}%
	\bibitem [{\citenamefont {Sun}\ \emph {et~al.}(2016)\citenamefont {Sun},
		\citenamefont {Matsuura}, \citenamefont {Ye}, \citenamefont {Mizukami},
		\citenamefont {Shimozawa}, \citenamefont {Matsubayashi}, \citenamefont
		{Yamashita}, \citenamefont {Watashige}, \citenamefont {Kasahara},
		\citenamefont {Matsuda}, \citenamefont {Yan}, \citenamefont {Sales},
		\citenamefont {Uwatoko}, \citenamefont {Cheng},\ and\ \citenamefont
		{Shibauchi}}]{Sun2016a}%
	\BibitemOpen
	\bibfield  {author} {\bibinfo {author} {\bibfnamefont {J.~P.}\ \bibnamefont
			{Sun}}, \bibinfo {author} {\bibfnamefont {K.}~\bibnamefont {Matsuura}},
		\bibinfo {author} {\bibfnamefont {G.~Z.}\ \bibnamefont {Ye}}, \bibinfo
		{author} {\bibfnamefont {Y.}~\bibnamefont {Mizukami}}, \bibinfo {author}
		{\bibfnamefont {M.}~\bibnamefont {Shimozawa}}, \bibinfo {author}
		{\bibfnamefont {K.}~\bibnamefont {Matsubayashi}}, \bibinfo {author}
		{\bibfnamefont {M.}~\bibnamefont {Yamashita}}, \bibinfo {author}
		{\bibfnamefont {T.}~\bibnamefont {Watashige}}, \bibinfo {author}
		{\bibfnamefont {S.}~\bibnamefont {Kasahara}}, \bibinfo {author}
		{\bibfnamefont {Y.}~\bibnamefont {Matsuda}}, \bibinfo {author} {\bibfnamefont
			{J.~Q.}\ \bibnamefont {Yan}}, \bibinfo {author} {\bibfnamefont {B.~C.}\
			\bibnamefont {Sales}}, \bibinfo {author} {\bibfnamefont {Y.}~\bibnamefont
			{Uwatoko}}, \bibinfo {author} {\bibfnamefont {J.~G.}\ \bibnamefont {Cheng}},
		\ and\ \bibinfo {author} {\bibfnamefont {T.}~\bibnamefont {Shibauchi}},\
	}\href {http://dx.doi.org/10.1038/ncomms12146} {\bibfield  {journal}
	{\bibinfo  {journal} {Nat. Commun.}\ }\textbf {\bibinfo {volume} {7}},\
	\bibinfo {pages} {12146} (\bibinfo {year} {2016})}\BibitemShut {NoStop}%
\bibitem [{\citenamefont {Sun}\ \emph {et~al.}(2017)\citenamefont {Sun},
	\citenamefont {Ye}, \citenamefont {Shahi}, \citenamefont {Yan}, \citenamefont
	{Matsuura}, \citenamefont {Kontani}, \citenamefont {Zhang}, \citenamefont
	{Zhou}, \citenamefont {Sales}, \citenamefont {Shibauchi}, \citenamefont
	{Uwatoko}, \citenamefont {Singh},\ and\ \citenamefont {Cheng}}]{Sun2017}%
\BibitemOpen
\bibfield  {author} {\bibinfo {author} {\bibfnamefont {J.~P.}\ \bibnamefont
		{Sun}}, \bibinfo {author} {\bibfnamefont {G.~Z.}\ \bibnamefont {Ye}},
	\bibinfo {author} {\bibfnamefont {P.}~\bibnamefont {Shahi}}, \bibinfo
	{author} {\bibfnamefont {J.-Q.}\ \bibnamefont {Yan}}, \bibinfo {author}
	{\bibfnamefont {K.}~\bibnamefont {Matsuura}}, \bibinfo {author}
	{\bibfnamefont {H.}~\bibnamefont {Kontani}}, \bibinfo {author} {\bibfnamefont
		{G.~M.}\ \bibnamefont {Zhang}}, \bibinfo {author} {\bibfnamefont
		{Q.}~\bibnamefont {Zhou}}, \bibinfo {author} {\bibfnamefont {B.~C.}\
		\bibnamefont {Sales}}, \bibinfo {author} {\bibfnamefont {T.}~\bibnamefont
		{Shibauchi}}, \bibinfo {author} {\bibfnamefont {Y.}~\bibnamefont {Uwatoko}},
	\bibinfo {author} {\bibfnamefont {D.~J.}\ \bibnamefont {Singh}}, \ and\
	\bibinfo {author} {\bibfnamefont {J.-G.}\ \bibnamefont {Cheng}},\ }\href
{\doibase 10.1103/PhysRevLett.118.147004} {\bibfield  {journal} {\bibinfo
		{journal} {Phys. Rev. Lett.}\ }\textbf {\bibinfo {volume} {118}},\ \bibinfo
	{pages} {147004} (\bibinfo {year} {2017})}\BibitemShut {NoStop}%
\bibitem [{\citenamefont {Kothapalli}\ \emph {et~al.}(2016)\citenamefont
	{Kothapalli}, \citenamefont {B\"ohmer}, \citenamefont {Jayasekara},
	\citenamefont {Ueland}, \citenamefont {Das}, \citenamefont {Sapkota},
	\citenamefont {Taufour}, \citenamefont {Xiao}, \citenamefont {Alp},
	\citenamefont {Bud'ko}, \citenamefont {Canfield}, \citenamefont {Kreyssig},\
	and\ \citenamefont {Goldman}}]{Kothapalli2016}%
\BibitemOpen
\bibfield  {author} {\bibinfo {author} {\bibfnamefont {K.}~\bibnamefont
		{Kothapalli}}, \bibinfo {author} {\bibfnamefont {A.~E.}\ \bibnamefont
		{B\"ohmer}}, \bibinfo {author} {\bibfnamefont {W.~T.}\ \bibnamefont
		{Jayasekara}}, \bibinfo {author} {\bibfnamefont {B.~G.}\ \bibnamefont
		{Ueland}}, \bibinfo {author} {\bibfnamefont {P.}~\bibnamefont {Das}},
	\bibinfo {author} {\bibfnamefont {A.}~\bibnamefont {Sapkota}}, \bibinfo
	{author} {\bibfnamefont {V.}~\bibnamefont {Taufour}}, \bibinfo {author}
	{\bibfnamefont {Y.}~\bibnamefont {Xiao}}, \bibinfo {author} {\bibfnamefont
		{E.}~\bibnamefont {Alp}}, \bibinfo {author} {\bibfnamefont {S.~L.}\
		\bibnamefont {Bud'ko}}, \bibinfo {author} {\bibfnamefont {P.~C.}\
		\bibnamefont {Canfield}}, \bibinfo {author} {\bibfnamefont {A.}~\bibnamefont
		{Kreyssig}}, \ and\ \bibinfo {author} {\bibfnamefont {A.~I.}\ \bibnamefont
		{Goldman}},\ }\href {http://dx.doi.org/10.1038/ncomms12728} {\bibfield
	{journal} {\bibinfo  {journal} {Nat. Commun.}\ }\textbf {\bibinfo {volume}
		{7}},\ \bibinfo {pages} {12728} (\bibinfo {year} {2016})}\BibitemShut
{NoStop}%
\bibitem [{\citenamefont {Glasbrenner}\ \emph {et~al.}(2015)\citenamefont
	{Glasbrenner}, \citenamefont {Mazin}, \citenamefont {Jeschke}, \citenamefont
	{Hirschfeld}, \citenamefont {Fernandes},\ and\ \citenamefont
	{Valenti}}]{Glasbrenner2015}%
\BibitemOpen
\bibfield  {author} {\bibinfo {author} {\bibfnamefont {J.~K.}\ \bibnamefont
		{Glasbrenner}}, \bibinfo {author} {\bibfnamefont {I.~I.}\ \bibnamefont
		{Mazin}}, \bibinfo {author} {\bibfnamefont {H.~O.}\ \bibnamefont {Jeschke}},
	\bibinfo {author} {\bibfnamefont {P.~J.}\ \bibnamefont {Hirschfeld}},
	\bibinfo {author} {\bibfnamefont {R.~M.}\ \bibnamefont {Fernandes}}, \ and\
	\bibinfo {author} {\bibfnamefont {R.}~\bibnamefont {Valenti}},\ }\href
{http://dx.doi.org/10.1038/nphys3434} {\bibfield  {journal} {\bibinfo
		{journal} {Nat Phys}\ }\textbf {\bibinfo {volume} {11}},\ \bibinfo {pages}
	{953} (\bibinfo {year} {2015})}\BibitemShut {NoStop}%
\bibitem [{\citenamefont {Yu}\ and\ \citenamefont {Si}(2015)}]{Yu2015}%
\BibitemOpen
\bibfield  {author} {\bibinfo {author} {\bibfnamefont {R.}~\bibnamefont
		{Yu}}\ and\ \bibinfo {author} {\bibfnamefont {Q.}~\bibnamefont {Si}},\ }\href
{\doibase 10.1103/PhysRevLett.115.116401} {\bibfield  {journal} {\bibinfo
		{journal} {Phys. Rev. Lett.}\ }\textbf {\bibinfo {volume} {115}},\ \bibinfo
	{pages} {116401} (\bibinfo {year} {2015})}\BibitemShut {NoStop}%
\bibitem [{\citenamefont {Wang}\ \emph
	{et~al.}(2015{\natexlab{a}})\citenamefont {Wang}, \citenamefont {Kivelson},\
	and\ \citenamefont {Lee}}]{Wang2015}%
\BibitemOpen
\bibfield  {author} {\bibinfo {author} {\bibfnamefont {F.}~\bibnamefont
		{Wang}}, \bibinfo {author} {\bibfnamefont {S.~A.}\ \bibnamefont {Kivelson}},
	\ and\ \bibinfo {author} {\bibfnamefont {D.-H.}\ \bibnamefont {Lee}},\ }\href
{http://dx.doi.org/10.1038/nphys3456} {\bibfield  {journal} {\bibinfo
		{journal} {Nat Phys}\ }\textbf {\bibinfo {volume} {11}},\ \bibinfo {pages}
	{959} (\bibinfo {year} {2015}{\natexlab{a}})}\BibitemShut {NoStop}%
\bibitem [{\citenamefont {Chubukov}\ \emph {et~al.}(2016)\citenamefont
	{Chubukov}, \citenamefont {Khodas},\ and\ \citenamefont
	{Fernandes}}]{Chubukov2016}%
\BibitemOpen
\bibfield  {author} {\bibinfo {author} {\bibfnamefont {A.~V.}\ \bibnamefont
		{Chubukov}}, \bibinfo {author} {\bibfnamefont {M.}~\bibnamefont {Khodas}}, \
	and\ \bibinfo {author} {\bibfnamefont {R.~M.}\ \bibnamefont {Fernandes}},\
}\href {\doibase 10.1103/PhysRevX.6.041045} {\bibfield  {journal} {\bibinfo
	{journal} {Phys. Rev. X}\ }\textbf {\bibinfo {volume} {6}},\ \bibinfo {pages}
{041045} (\bibinfo {year} {2016})}\BibitemShut {NoStop}%
\bibitem [{\citenamefont {Miyoshi}(2014)}]{Miyoshi2014}%
\BibitemOpen
\bibfield  {author} {\bibinfo {author} {\bibfnamefont {K.}~\bibnamefont
		{Miyoshi}},\ }\href {\doibase 10.7566/jpsj.83.013702} {\bibfield  {journal}
	{\bibinfo  {journal} {J. Phys. Soc. Jpn.}\ }\textbf {\bibinfo {volume}
		{83}},\ \bibinfo {pages} {013702} (\bibinfo {year} {2014})}\BibitemShut
{NoStop}%
\bibitem [{\citenamefont {Ge}\ \emph {et~al.}(2015)\citenamefont {Ge},
	\citenamefont {Liu}, \citenamefont {Liu}, \citenamefont {Gao}, \citenamefont
	{Qian}, \citenamefont {Xue}, \citenamefont {Liu},\ and\ \citenamefont
	{Jia}}]{Ge2015}%
\BibitemOpen
\bibfield  {author} {\bibinfo {author} {\bibfnamefont {J.-F.}\ \bibnamefont
		{Ge}}, \bibinfo {author} {\bibfnamefont {Z.-L.}\ \bibnamefont {Liu}},
	\bibinfo {author} {\bibfnamefont {C.}~\bibnamefont {Liu}}, \bibinfo {author}
	{\bibfnamefont {C.-L.}\ \bibnamefont {Gao}}, \bibinfo {author} {\bibfnamefont
		{D.}~\bibnamefont {Qian}}, \bibinfo {author} {\bibfnamefont {Q.-K.}\
		\bibnamefont {Xue}}, \bibinfo {author} {\bibfnamefont {Y.}~\bibnamefont
		{Liu}}, \ and\ \bibinfo {author} {\bibfnamefont {J.-F.}\ \bibnamefont
		{Jia}},\ }\href {http://dx.doi.org/10.1038/nmat4153} {\bibfield  {journal}
	{\bibinfo  {journal} {Nat Mater}\ }\textbf {\bibinfo {volume} {14}},\
	\bibinfo {pages} {285} (\bibinfo {year} {2015})}\BibitemShut {NoStop}%
\bibitem [{\citenamefont {Mizuguchi}\ \emph {et~al.}(2009)\citenamefont
	{Mizuguchi}, \citenamefont {Tomioka}, \citenamefont {Tsuda}, \citenamefont
	{Yamaguchi},\ and\ \citenamefont {Takano}}]{Mizuguchi2009}%
\BibitemOpen
\bibfield  {author} {\bibinfo {author} {\bibfnamefont {Y.}~\bibnamefont
		{Mizuguchi}}, \bibinfo {author} {\bibfnamefont {F.}~\bibnamefont {Tomioka}},
	\bibinfo {author} {\bibfnamefont {S.}~\bibnamefont {Tsuda}}, \bibinfo
	{author} {\bibfnamefont {T.}~\bibnamefont {Yamaguchi}}, \ and\ \bibinfo
	{author} {\bibfnamefont {Y.}~\bibnamefont {Takano}},\ }\href {\doibase
	10.1143/JPSJ.78.074712} {\bibfield  {journal} {\bibinfo  {journal} {J. Phys.
			Soc. Jpn.}\ }\textbf {\bibinfo {volume} {78}},\ \bibinfo {pages} {074712}
	(\bibinfo {year} {2009})}\BibitemShut {NoStop}%
\bibitem [{\citenamefont {Watson}\ \emph {et~al.}(2015)\citenamefont {Watson},
	\citenamefont {Kim}, \citenamefont {Haghighirad}, \citenamefont {Blake},
	\citenamefont {Davies}, \citenamefont {Hoesch}, \citenamefont {Wolf},\ and\
	\citenamefont {Coldea}}]{Watson2015II}%
\BibitemOpen
\bibfield  {author} {\bibinfo {author} {\bibfnamefont {M.~D.}\ \bibnamefont
		{Watson}}, \bibinfo {author} {\bibfnamefont {T.~K.}\ \bibnamefont {Kim}},
	\bibinfo {author} {\bibfnamefont {A.~A.}\ \bibnamefont {Haghighirad}},
	\bibinfo {author} {\bibfnamefont {S.~F.}\ \bibnamefont {Blake}}, \bibinfo
	{author} {\bibfnamefont {N.~R.}\ \bibnamefont {Davies}}, \bibinfo {author}
	{\bibfnamefont {M.}~\bibnamefont {Hoesch}}, \bibinfo {author} {\bibfnamefont
		{T.}~\bibnamefont {Wolf}}, \ and\ \bibinfo {author} {\bibfnamefont {A.~I.}\
		\bibnamefont {Coldea}},\ }\href {\doibase 10.1103/PhysRevB.92.121108}
{\bibfield  {journal} {\bibinfo  {journal} {Phys. Rev. B}\ }\textbf {\bibinfo
		{volume} {92}},\ \bibinfo {pages} {121108} (\bibinfo {year}
	{2015})}\BibitemShut {NoStop}%
\bibitem [{\citenamefont {Coldea}\ \emph {et~al.}(2016)\citenamefont {Coldea},
	\citenamefont {Blake}, \citenamefont {Kasahara}, \citenamefont {Haghighirad},
	\citenamefont {Watson}, \citenamefont {Knafo}, \citenamefont {Choi},
	\citenamefont {McCollam}, \citenamefont {Reiss}, \citenamefont {Yamashita},
	\citenamefont {Bruma}, \citenamefont {Speller}, \citenamefont {Matsuda},
	\citenamefont {Wolf}, \citenamefont {Shibauchi},\ and\ \citenamefont
	{Schofield}}]{Coldea2016}%
\BibitemOpen
\bibfield  {author} {\bibinfo {author} {\bibfnamefont {A.~I.}\ \bibnamefont
		{Coldea}}, \bibinfo {author} {\bibfnamefont {S.~F.}\ \bibnamefont {Blake}},
	\bibinfo {author} {\bibfnamefont {S.}~\bibnamefont {Kasahara}}, \bibinfo
	{author} {\bibfnamefont {A.~A.}\ \bibnamefont {Haghighirad}}, \bibinfo
	{author} {\bibfnamefont {M.~D.}\ \bibnamefont {Watson}}, \bibinfo {author}
	{\bibfnamefont {W.}~\bibnamefont {Knafo}}, \bibinfo {author} {\bibfnamefont
		{E.~S.}\ \bibnamefont {Choi}}, \bibinfo {author} {\bibfnamefont
		{A.}~\bibnamefont {McCollam}}, \bibinfo {author} {\bibfnamefont
		{P.}~\bibnamefont {Reiss}}, \bibinfo {author} {\bibfnamefont
		{T.}~\bibnamefont {Yamashita}}, \bibinfo {author} {\bibfnamefont
		{M.}~\bibnamefont {Bruma}}, \bibinfo {author} {\bibfnamefont
		{S.}~\bibnamefont {Speller}}, \bibinfo {author} {\bibfnamefont
		{Y.}~\bibnamefont {Matsuda}}, \bibinfo {author} {\bibfnamefont
		{T.}~\bibnamefont {Wolf}}, \bibinfo {author} {\bibfnamefont {T.}~\bibnamefont
		{Shibauchi}}, \ and\ \bibinfo {author} {\bibfnamefont {A.~J.}\ \bibnamefont
		{Schofield}},\ }\href {https://arxiv.org/abs/1611.07424} {\bibfield
	{journal} {\bibinfo  {journal} {arXiv:1611.07424}\ } (\bibinfo {year}
	{2016})}\BibitemShut {NoStop}%
\bibitem [{\citenamefont {Hosoi}\ \emph {et~al.}(2016)\citenamefont {Hosoi},
	\citenamefont {Matsuura}, \citenamefont {Ishida}, \citenamefont {Wang},
	\citenamefont {Mizukami}, \citenamefont {Watashige}, \citenamefont
	{Kasahara}, \citenamefont {Matsuda},\ and\ \citenamefont
	{Shibauchi}}]{Hosoi2016}%
\BibitemOpen
\bibfield  {author} {\bibinfo {author} {\bibfnamefont {S.}~\bibnamefont
		{Hosoi}}, \bibinfo {author} {\bibfnamefont {K.}~\bibnamefont {Matsuura}},
	\bibinfo {author} {\bibfnamefont {K.}~\bibnamefont {Ishida}}, \bibinfo
	{author} {\bibfnamefont {H.}~\bibnamefont {Wang}}, \bibinfo {author}
	{\bibfnamefont {Y.}~\bibnamefont {Mizukami}}, \bibinfo {author}
	{\bibfnamefont {T.}~\bibnamefont {Watashige}}, \bibinfo {author}
	{\bibfnamefont {S.}~\bibnamefont {Kasahara}}, \bibinfo {author}
	{\bibfnamefont {Y.}~\bibnamefont {Matsuda}}, \ and\ \bibinfo {author}
	{\bibfnamefont {T.}~\bibnamefont {Shibauchi}},\ }\href {\doibase
	10.1073/pnas.1605806113} {\bibfield  {journal} {\bibinfo  {journal} {Proc.
			Natl. Acad. Sci. U.S.A.}\ }\textbf {\bibinfo {volume} {113}},\ \bibinfo
	{pages} {8139} (\bibinfo {year} {2016})}\BibitemShut {NoStop}%
\bibitem [{\citenamefont {Ovchenkov}\ \emph {et~al.}(2016)\citenamefont
	{Ovchenkov}, \citenamefont {Chareev}, \citenamefont {Presnov}, \citenamefont
	{Volkova},\ and\ \citenamefont {Vasiliev}}]{Ovchenkov2016}%
\BibitemOpen
\bibfield  {author} {\bibinfo {author} {\bibfnamefont {Y.~A.}\ \bibnamefont
		{Ovchenkov}}, \bibinfo {author} {\bibfnamefont {D.~A.}\ \bibnamefont
		{Chareev}}, \bibinfo {author} {\bibfnamefont {D.~E.}\ \bibnamefont
		{Presnov}}, \bibinfo {author} {\bibfnamefont {O.~S.}\ \bibnamefont
		{Volkova}}, \ and\ \bibinfo {author} {\bibfnamefont {A.~N.}\ \bibnamefont
		{Vasiliev}},\ }\href {http://dx.doi.org/10.1007/s10909-016-1593-x} {\bibfield
	{journal} {\bibinfo  {journal} {J. Low Temp. Phys.}\ }\textbf {\bibinfo
		{volume} {185}},\ \bibinfo {pages} {467} (\bibinfo {year}
	{2016})}\BibitemShut {NoStop}%
\bibitem [{\citenamefont {B\"ohmer}\ \emph {et~al.}(2016)\citenamefont
	{B\"ohmer}, \citenamefont {Taufour}, \citenamefont {Straszheim},
	\citenamefont {Wolf},\ and\ \citenamefont {Canfield}}]{Boehmer2016}%
\BibitemOpen
\bibfield  {author} {\bibinfo {author} {\bibfnamefont {A.~E.}\ \bibnamefont
		{B\"ohmer}}, \bibinfo {author} {\bibfnamefont {V.}~\bibnamefont {Taufour}},
	\bibinfo {author} {\bibfnamefont {W.~E.}\ \bibnamefont {Straszheim}},
	\bibinfo {author} {\bibfnamefont {T.}~\bibnamefont {Wolf}}, \ and\ \bibinfo
	{author} {\bibfnamefont {P.~C.}\ \bibnamefont {Canfield}},\ }\href {\doibase
	10.1103/PhysRevB.94.024526} {\bibfield  {journal} {\bibinfo  {journal} {Phys.
			Rev. B}\ }\textbf {\bibinfo {volume} {94}},\ \bibinfo {pages} {024526}
	(\bibinfo {year} {2016})}\BibitemShut {NoStop}%
\bibitem [{\citenamefont {Tanatar}\ \emph {et~al.}(2009)\citenamefont
	{Tanatar}, \citenamefont {Ni}, \citenamefont {Samolyuk}, \citenamefont
	{Bud'ko}, \citenamefont {Canfield},\ and\ \citenamefont
	{Prozorov}}]{Tanatar2009}%
\BibitemOpen
\bibfield  {author} {\bibinfo {author} {\bibfnamefont {M.~A.}\ \bibnamefont
		{Tanatar}}, \bibinfo {author} {\bibfnamefont {N.}~\bibnamefont {Ni}},
	\bibinfo {author} {\bibfnamefont {G.~D.}\ \bibnamefont {Samolyuk}}, \bibinfo
	{author} {\bibfnamefont {S.~L.}\ \bibnamefont {Bud'ko}}, \bibinfo {author}
	{\bibfnamefont {P.~C.}\ \bibnamefont {Canfield}}, \ and\ \bibinfo {author}
	{\bibfnamefont {R.}~\bibnamefont {Prozorov}},\ }\href {\doibase
	10.1103/PhysRevB.79.134528} {\bibfield  {journal} {\bibinfo  {journal} {Phys.
			Rev. B}\ }\textbf {\bibinfo {volume} {79}},\ \bibinfo {pages} {134528}
	(\bibinfo {year} {2009})}\BibitemShut {NoStop}%
\bibitem [{\citenamefont {Tanatar}\ \emph {et~al.}(2016)\citenamefont
	{Tanatar}, \citenamefont {B\"ohmer}, \citenamefont {Timmons}, \citenamefont
	{Sch\"utt}, \citenamefont {Drachuck}, \citenamefont {Taufour}, \citenamefont
	{Kothapalli}, \citenamefont {Kreyssig}, \citenamefont {Bud'ko}, \citenamefont
	{Canfield}, \citenamefont {Fernandes},\ and\ \citenamefont
	{Prozorov}}]{Tanatar2016}%
\BibitemOpen
\bibfield  {author} {\bibinfo {author} {\bibfnamefont {M.~A.}\ \bibnamefont
		{Tanatar}}, \bibinfo {author} {\bibfnamefont {A.~E.}\ \bibnamefont
		{B\"ohmer}}, \bibinfo {author} {\bibfnamefont {E.~I.}\ \bibnamefont
		{Timmons}}, \bibinfo {author} {\bibfnamefont {M.}~\bibnamefont {Sch\"utt}},
	\bibinfo {author} {\bibfnamefont {G.}~\bibnamefont {Drachuck}}, \bibinfo
	{author} {\bibfnamefont {V.}~\bibnamefont {Taufour}}, \bibinfo {author}
	{\bibfnamefont {K.}~\bibnamefont {Kothapalli}}, \bibinfo {author}
	{\bibfnamefont {A.}~\bibnamefont {Kreyssig}}, \bibinfo {author}
	{\bibfnamefont {S.~L.}\ \bibnamefont {Bud'ko}}, \bibinfo {author}
	{\bibfnamefont {P.~C.}\ \bibnamefont {Canfield}}, \bibinfo {author}
	{\bibfnamefont {R.~M.}\ \bibnamefont {Fernandes}}, \ and\ \bibinfo {author}
	{\bibfnamefont {R.}~\bibnamefont {Prozorov}},\ }\href {\doibase
	10.1103/PhysRevLett.117.127001} {\bibfield  {journal} {\bibinfo  {journal}
		{Phys. Rev. Lett.}\ }\textbf {\bibinfo {volume} {117}},\ \bibinfo {pages}
	{127001} (\bibinfo {year} {2016})}\BibitemShut {NoStop}%
\bibitem [{\citenamefont {Bud'ko}\ \emph {et~al.}(1984)\citenamefont {Bud'ko},
	\citenamefont {Voronovskii}, \citenamefont {Gapotchenko},\ and\ \citenamefont
	{ltskevich}}]{Budko1984}%
\BibitemOpen
\bibfield  {author} {\bibinfo {author} {\bibfnamefont {S.}~\bibnamefont
		{Bud'ko}}, \bibinfo {author} {\bibfnamefont {A.}~\bibnamefont {Voronovskii}},
	\bibinfo {author} {\bibfnamefont {A.}~\bibnamefont {Gapotchenko}}, \ and\
	\bibinfo {author} {\bibfnamefont {E.}~\bibnamefont {ltskevich}},\ }\href
{http://www.jetp.ac.ru/cgi-bin/e/index/e/59/2/p454?a=list} {\bibfield
	{journal} {\bibinfo  {journal} {Zh. Eksp. Teor. Fiz. 86, 778}\ } (\bibinfo
	{year} {1984})}\BibitemShut {NoStop}%
\bibitem [{\citenamefont {Bireckoven}\ and\ \citenamefont
	{Wittig}(1988)}]{Bireckoven1988}%
\BibitemOpen
\bibfield  {author} {\bibinfo {author} {\bibfnamefont {B.}~\bibnamefont
		{Bireckoven}}\ and\ \bibinfo {author} {\bibfnamefont {J.}~\bibnamefont
		{Wittig}},\ }\href {http://stacks.iop.org/0022-3735/21/i=9/a=004} {\bibfield
	{journal} {\bibinfo  {journal} {Meas. Sci. Technol.}\ }\textbf {\bibinfo
		{volume} {21}},\ \bibinfo {pages} {841} (\bibinfo {year} {1988})}\BibitemShut
{NoStop}%
\bibitem [{\citenamefont {Kim}\ \emph {et~al.}(2011)\citenamefont {Kim},
	\citenamefont {Torikachvili}, \citenamefont {Colombier}, \citenamefont
	{Thaler}, \citenamefont {Bud'ko},\ and\ \citenamefont {Canfield}}]{Kim2011}%
\BibitemOpen
\bibfield  {author} {\bibinfo {author} {\bibfnamefont {S.~K.}\ \bibnamefont
		{Kim}}, \bibinfo {author} {\bibfnamefont {M.~S.}\ \bibnamefont
		{Torikachvili}}, \bibinfo {author} {\bibfnamefont {E.}~\bibnamefont
		{Colombier}}, \bibinfo {author} {\bibfnamefont {A.}~\bibnamefont {Thaler}},
	\bibinfo {author} {\bibfnamefont {S.~L.}\ \bibnamefont {Bud'ko}}, \ and\
	\bibinfo {author} {\bibfnamefont {P.~C.}\ \bibnamefont {Canfield}},\ }\href
{\doibase 10.1103/PhysRevB.84.134525} {\bibfield  {journal} {\bibinfo
		{journal} {Phys. Rev. B}\ }\textbf {\bibinfo {volume} {84}},\ \bibinfo
	{pages} {134525} (\bibinfo {year} {2011})}\BibitemShut {NoStop}%
\bibitem [{\citenamefont {Torikachvili}\ \emph {et~al.}(2015)\citenamefont
	{Torikachvili}, \citenamefont {Kim}, \citenamefont {Colombier}, \citenamefont
	{Bud’ko},\ and\ \citenamefont {Canfield}}]{Torikachvili2015}%
\BibitemOpen
\bibfield  {author} {\bibinfo {author} {\bibfnamefont {M.~S.}\ \bibnamefont
		{Torikachvili}}, \bibinfo {author} {\bibfnamefont {S.~K.}\ \bibnamefont
		{Kim}}, \bibinfo {author} {\bibfnamefont {E.}~\bibnamefont {Colombier}},
	\bibinfo {author} {\bibfnamefont {S.~L.}\ \bibnamefont {Bud’ko}}, \ and\
	\bibinfo {author} {\bibfnamefont {P.~C.}\ \bibnamefont {Canfield}},\ }\href
{http://aip.scitation.org/doi/abs/10.1063/1.4937478} {\bibfield  {journal}
	{\bibinfo  {journal} {Rev. Sci. Instrum.}\ } (\bibinfo {year}
	{2015})}\BibitemShut {NoStop}%
\bibitem [{\citenamefont {Terashima}\ \emph {et~al.}(2016)\citenamefont
	{Terashima}, \citenamefont {Kikugawa}, \citenamefont {Kiswandhi},
	\citenamefont {Graf}, \citenamefont {Choi}, \citenamefont {Brooks},
	\citenamefont {Kasahara}, \citenamefont {Watashige}, \citenamefont {Matsuda},
	\citenamefont {Shibauchi}, \citenamefont {Wolf}, \citenamefont {B\"ohmer},
	\citenamefont {Hardy}, \citenamefont {Meingast}, \citenamefont
	{L\"ohneysen},\ and\ \citenamefont {Uji}}]{Terashima2016}%
\BibitemOpen
\bibfield  {author} {\bibinfo {author} {\bibfnamefont {T.}~\bibnamefont
		{Terashima}}, \bibinfo {author} {\bibfnamefont {N.}~\bibnamefont {Kikugawa}},
	\bibinfo {author} {\bibfnamefont {A.}~\bibnamefont {Kiswandhi}}, \bibinfo
	{author} {\bibfnamefont {D.}~\bibnamefont {Graf}}, \bibinfo {author}
	{\bibfnamefont {E.-S.}\ \bibnamefont {Choi}}, \bibinfo {author}
	{\bibfnamefont {J.~S.}\ \bibnamefont {Brooks}}, \bibinfo {author}
	{\bibfnamefont {S.}~\bibnamefont {Kasahara}}, \bibinfo {author}
	{\bibfnamefont {T.}~\bibnamefont {Watashige}}, \bibinfo {author}
	{\bibfnamefont {Y.}~\bibnamefont {Matsuda}}, \bibinfo {author} {\bibfnamefont
		{T.}~\bibnamefont {Shibauchi}}, \bibinfo {author} {\bibfnamefont
		{T.}~\bibnamefont {Wolf}}, \bibinfo {author} {\bibfnamefont {A.~E.}\
		\bibnamefont {B\"ohmer}}, \bibinfo {author} {\bibfnamefont {F.}~\bibnamefont
		{Hardy}}, \bibinfo {author} {\bibfnamefont {C.}~\bibnamefont {Meingast}},
	\bibinfo {author} {\bibfnamefont {H.~v.}\ \bibnamefont {L\"ohneysen}}, \ and\
	\bibinfo {author} {\bibfnamefont {S.}~\bibnamefont {Uji}},\ }\href {\doibase
	10.1103/PhysRevB.93.094505} {\bibfield  {journal} {\bibinfo  {journal} {Phys.
			Rev. B}\ }\textbf {\bibinfo {volume} {93}},\ \bibinfo {pages} {094505}
	(\bibinfo {year} {2016})}\BibitemShut {NoStop}%
\bibitem [{\citenamefont {Lederer}\ \emph {et~al.}(2015)\citenamefont
	{Lederer}, \citenamefont {Schattner}, \citenamefont {Berg},\ and\
	\citenamefont {Kivelson}}]{Lederer2015}%
\BibitemOpen
\bibfield  {author} {\bibinfo {author} {\bibfnamefont {S.}~\bibnamefont
		{Lederer}}, \bibinfo {author} {\bibfnamefont {Y.}~\bibnamefont {Schattner}},
	\bibinfo {author} {\bibfnamefont {E.}~\bibnamefont {Berg}}, \ and\ \bibinfo
	{author} {\bibfnamefont {S.~A.}\ \bibnamefont {Kivelson}},\ }\href {\doibase
	10.1103/PhysRevLett.114.097001} {\bibfield  {journal} {\bibinfo  {journal}
		{Phys. Rev. Lett.}\ }\textbf {\bibinfo {volume} {114}},\ \bibinfo {pages}
	{097001} (\bibinfo {year} {2015})}\BibitemShut {NoStop}%
\bibitem [{\citenamefont {{Labat}}\ and\ \citenamefont
	{{Paul}}(2017)}]{Labat2017}%
\BibitemOpen
\bibfield  {author} {\bibinfo {author} {\bibfnamefont {D.}~\bibnamefont
		{{Labat}}}\ and\ \bibinfo {author} {\bibfnamefont {I.}~\bibnamefont
		{{Paul}}},\ }\href {https://arxiv.org/abs/1703.04146} {\bibfield  {journal}
	{\bibinfo  {journal} {arXiv:1703.04146}\ ,\ \bibinfo {pages} {1703.04146}}
	(\bibinfo {year} {2017})}\BibitemShut {NoStop}%
\bibitem [{\citenamefont {Fernandes}\ and\ \citenamefont
	{Schmalian}(2012)}]{Fernandes2012}%
\BibitemOpen
\bibfield  {author} {\bibinfo {author} {\bibfnamefont {R.~M.}\ \bibnamefont
		{Fernandes}}\ and\ \bibinfo {author} {\bibfnamefont {J.}~\bibnamefont
		{Schmalian}},\ }\href {http://stacks.iop.org/0953-2048/25/i=8/a=084005}
{\bibfield  {journal} {\bibinfo  {journal} {Supercond. Sci. Technol.}\
	}\textbf {\bibinfo {volume} {25}},\ \bibinfo {pages} {084005} (\bibinfo
	{year} {2012})}\BibitemShut {NoStop}%
\bibitem [{\citenamefont {Wang}\ \emph {et~al.}(2016)\citenamefont {Wang},
	\citenamefont {Hu},\ and\ \citenamefont {Nevidomskyy}}]{Wang2016}%
\BibitemOpen
\bibfield  {author} {\bibinfo {author} {\bibfnamefont {Z.}~\bibnamefont
		{Wang}}, \bibinfo {author} {\bibfnamefont {W.-J.}\ \bibnamefont {Hu}}, \ and\
	\bibinfo {author} {\bibfnamefont {A.~H.}\ \bibnamefont {Nevidomskyy}},\
}\href {\doibase 10.1103/PhysRevLett.116.247203} {\bibfield  {journal}
{\bibinfo  {journal} {Phys. Rev. Lett.}\ }\textbf {\bibinfo {volume} {116}},\
\bibinfo {pages} {247203} (\bibinfo {year} {2016})}\BibitemShut {NoStop}%
\bibitem [{\citenamefont {Wang}\ \emph
	{et~al.}(2015{\natexlab{b}})\citenamefont {Wang}, \citenamefont {Kivelson},\
	and\ \citenamefont {Lee}}]{Wang2015a}%
\BibitemOpen
\bibfield  {author} {\bibinfo {author} {\bibfnamefont {F.}~\bibnamefont
		{Wang}}, \bibinfo {author} {\bibfnamefont {S.~A.}\ \bibnamefont {Kivelson}},
	\ and\ \bibinfo {author} {\bibfnamefont {D.-H.}\ \bibnamefont {Lee}},\ }\href
{http://dx.doi.org/10.1038/nphys3456} {\bibfield  {journal} {\bibinfo
		{journal} {Nat Phys}\ }\textbf {\bibinfo {volume} {11}},\ \bibinfo {pages}
	{959} (\bibinfo {year} {2015}{\natexlab{b}})}\BibitemShut {NoStop}%
\bibitem [{\citenamefont {Klintberg}\ \emph {et~al.}(2010)\citenamefont
	{Klintberg}, \citenamefont {Goh}, \citenamefont {Kasahara}, \citenamefont
	{Nakai}, \citenamefont {Ishida}, \citenamefont {Sutherland}, \citenamefont
	{Shibauchi}, \citenamefont {Matsuda},\ and\ \citenamefont
	{Terashima}}]{Klintberg2010}%
\BibitemOpen
\bibfield  {author} {\bibinfo {author} {\bibfnamefont {L.~E.}\ \bibnamefont
		{Klintberg}}, \bibinfo {author} {\bibfnamefont {S.~K.}\ \bibnamefont {Goh}},
	\bibinfo {author} {\bibfnamefont {S.}~\bibnamefont {Kasahara}}, \bibinfo
	{author} {\bibfnamefont {Y.}~\bibnamefont {Nakai}}, \bibinfo {author}
	{\bibfnamefont {K.}~\bibnamefont {Ishida}}, \bibinfo {author} {\bibfnamefont
		{M.}~\bibnamefont {Sutherland}}, \bibinfo {author} {\bibfnamefont
		{T.}~\bibnamefont {Shibauchi}}, \bibinfo {author} {\bibfnamefont
		{Y.}~\bibnamefont {Matsuda}}, \ and\ \bibinfo {author} {\bibfnamefont
		{T.}~\bibnamefont {Terashima}},\ }\href {\doibase 10.1143/JPSJ.79.123706}
{\bibfield  {journal} {\bibinfo  {journal} {J. Phys. Soc. Jpn.}\ }\textbf
	{\bibinfo {volume} {79}},\ \bibinfo {pages} {123706} (\bibinfo {year}
	{2010})}\BibitemShut {NoStop}%
\bibitem [{\citenamefont {Jiang}\ \emph {et~al.}(2009)\citenamefont {Jiang},
	\citenamefont {Xing}, \citenamefont {Xuan}, \citenamefont {Wang},
	\citenamefont {Ren}, \citenamefont {Feng}, \citenamefont {Dai}, \citenamefont
	{Xu},\ and\ \citenamefont {Cao}}]{Jiang2009}%
\BibitemOpen
\bibfield  {author} {\bibinfo {author} {\bibfnamefont {S.}~\bibnamefont
		{Jiang}}, \bibinfo {author} {\bibfnamefont {H.}~\bibnamefont {Xing}},
	\bibinfo {author} {\bibfnamefont {G.}~\bibnamefont {Xuan}}, \bibinfo {author}
	{\bibfnamefont {C.}~\bibnamefont {Wang}}, \bibinfo {author} {\bibfnamefont
		{Z.}~\bibnamefont {Ren}}, \bibinfo {author} {\bibfnamefont {C.}~\bibnamefont
		{Feng}}, \bibinfo {author} {\bibfnamefont {J.}~\bibnamefont {Dai}}, \bibinfo
	{author} {\bibfnamefont {Z.}~\bibnamefont {Xu}}, \ and\ \bibinfo {author}
	{\bibfnamefont {G.}~\bibnamefont {Cao}},\ }\href
{http://stacks.iop.org/0953-8984/21/i=38/a=382203} {\bibfield  {journal}
	{\bibinfo  {journal} {J. Phys. Condens. Matter}\ }\textbf {\bibinfo {volume}
		{21}},\ \bibinfo {pages} {382203} (\bibinfo {year} {2009})}\BibitemShut
{NoStop}%
\bibitem [{\citenamefont {Colombier}\ \emph {et~al.}(2009)\citenamefont
	{Colombier}, \citenamefont {Bud'ko}, \citenamefont {Ni},\ and\ \citenamefont
	{Canfield}}]{Colombier2009}%
\BibitemOpen
\bibfield  {author} {\bibinfo {author} {\bibfnamefont {E.}~\bibnamefont
		{Colombier}}, \bibinfo {author} {\bibfnamefont {S.~L.}\ \bibnamefont
		{Bud'ko}}, \bibinfo {author} {\bibfnamefont {N.}~\bibnamefont {Ni}}, \ and\
	\bibinfo {author} {\bibfnamefont {P.~C.}\ \bibnamefont {Canfield}},\ }\href
{\doibase 10.1103/PhysRevB.79.224518} {\bibfield  {journal} {\bibinfo
		{journal} {Phys. Rev. B}\ }\textbf {\bibinfo {volume} {79}},\ \bibinfo
	{pages} {224518} (\bibinfo {year} {2009})}\BibitemShut {NoStop}%
\bibitem [{\citenamefont {Thaler}\ \emph {et~al.}(2010)\citenamefont {Thaler},
	\citenamefont {Ni}, \citenamefont {Kracher}, \citenamefont {Yan},
	\citenamefont {Bud'ko},\ and\ \citenamefont {Canfield}}]{Thaler2010}%
\BibitemOpen
\bibfield  {author} {\bibinfo {author} {\bibfnamefont {A.}~\bibnamefont
		{Thaler}}, \bibinfo {author} {\bibfnamefont {N.}~\bibnamefont {Ni}}, \bibinfo
	{author} {\bibfnamefont {A.}~\bibnamefont {Kracher}}, \bibinfo {author}
	{\bibfnamefont {J.~Q.}\ \bibnamefont {Yan}}, \bibinfo {author} {\bibfnamefont
		{S.~L.}\ \bibnamefont {Bud'ko}}, \ and\ \bibinfo {author} {\bibfnamefont
		{P.~C.}\ \bibnamefont {Canfield}},\ }\href {\doibase
	10.1103/PhysRevB.82.014534} {\bibfield  {journal} {\bibinfo  {journal} {Phys.
			Rev. B}\ }\textbf {\bibinfo {volume} {82}},\ \bibinfo {pages} {014534}
	(\bibinfo {year} {2010})}\BibitemShut {NoStop}%
\bibitem [{\citenamefont {Taufour}\ \emph {et~al.}(2014)\citenamefont
	{Taufour}, \citenamefont {Foroozani}, \citenamefont {Tanatar}, \citenamefont
	{Lim}, \citenamefont {Kaluarachchi}, \citenamefont {Kim}, \citenamefont
	{Liu}, \citenamefont {Lograsso}, \citenamefont {Kogan}, \citenamefont
	{Prozorov}, \citenamefont {Bud'ko}, \citenamefont {Schilling},\ and\
	\citenamefont {Canfield}}]{Taufour2014}%
\BibitemOpen
\bibfield  {author} {\bibinfo {author} {\bibfnamefont {V.}~\bibnamefont
		{Taufour}}, \bibinfo {author} {\bibfnamefont {N.}~\bibnamefont {Foroozani}},
	\bibinfo {author} {\bibfnamefont {M.~A.}\ \bibnamefont {Tanatar}}, \bibinfo
	{author} {\bibfnamefont {J.}~\bibnamefont {Lim}}, \bibinfo {author}
	{\bibfnamefont {U.}~\bibnamefont {Kaluarachchi}}, \bibinfo {author}
	{\bibfnamefont {S.~K.}\ \bibnamefont {Kim}}, \bibinfo {author} {\bibfnamefont
		{Y.}~\bibnamefont {Liu}}, \bibinfo {author} {\bibfnamefont {T.~A.}\
		\bibnamefont {Lograsso}}, \bibinfo {author} {\bibfnamefont {V.~G.}\
		\bibnamefont {Kogan}}, \bibinfo {author} {\bibfnamefont {R.}~\bibnamefont
		{Prozorov}}, \bibinfo {author} {\bibfnamefont {S.~L.}\ \bibnamefont
		{Bud'ko}}, \bibinfo {author} {\bibfnamefont {J.~S.}\ \bibnamefont
		{Schilling}}, \ and\ \bibinfo {author} {\bibfnamefont {P.~C.}\ \bibnamefont
		{Canfield}},\ }\href {\doibase 10.1103/PhysRevB.89.220509} {\bibfield
	{journal} {\bibinfo  {journal} {Phys. Rev. B}\ }\textbf {\bibinfo {volume}
		{89}},\ \bibinfo {pages} {220509} (\bibinfo {year} {2014})}\BibitemShut
{NoStop}%
\bibitem [{\citenamefont {Kogan}\ and\ \citenamefont
	{Prozorov}(2012)}]{Kogan2012}%
\BibitemOpen
\bibfield  {author} {\bibinfo {author} {\bibfnamefont {V.~G.}\ \bibnamefont
		{Kogan}}\ and\ \bibinfo {author} {\bibfnamefont {R.}~\bibnamefont
		{Prozorov}},\ }\href {http://stacks.iop.org/0034-4885/75/i=11/a=114502}
{\bibfield  {journal} {\bibinfo  {journal} {Rep. Prog. Phys.}\ }\textbf
	{\bibinfo {volume} {75}},\ \bibinfo {pages} {114502} (\bibinfo {year}
	{2012})}\BibitemShut {NoStop}%
\bibitem [{\citenamefont {Kogan}\ and\ \citenamefont
	{Prozorov}(2014)}]{Kogan2014}%
\BibitemOpen
\bibfield  {author} {\bibinfo {author} {\bibfnamefont {V.~G.}\ \bibnamefont
		{Kogan}}\ and\ \bibinfo {author} {\bibfnamefont {R.}~\bibnamefont
		{Prozorov}},\ }\href {\doibase 10.1103/PhysRevB.90.180502} {\bibfield
	{journal} {\bibinfo  {journal} {Phys. Rev. B}\ }\textbf {\bibinfo {volume}
		{90}},\ \bibinfo {pages} {180502} (\bibinfo {year} {2014})}\BibitemShut
{NoStop}%
\bibitem [{\citenamefont {Matsuura}\ \emph {et~al.}(2017)\citenamefont
	{Matsuura}, \citenamefont {Mizukami}, \citenamefont {Arai}, \citenamefont
	{Sugimura}, \citenamefont {Maejima}, \citenamefont {Machida}, \citenamefont
	{Watanuki}, \citenamefont {Fukuda}, \citenamefont {Yajima}, \citenamefont
	{Hiroi}, \citenamefont {Yip}, \citenamefont {Chan}, \citenamefont {Niu},
	\citenamefont {Hosoi}, \citenamefont {Ishida}, \citenamefont {Watashige},
	\citenamefont {Kasahara}, \citenamefont {Cheng}, \citenamefont {Goh},
	\citenamefont {Matsuda}, \citenamefont {Uwatoko},\ and\ \citenamefont
	{Shibauchi}}]{Matsuura2017}%
\BibitemOpen
\bibfield  {author} {\bibinfo {author} {\bibfnamefont {K.}~\bibnamefont
		{Matsuura}}, \bibinfo {author} {\bibfnamefont {Y.}~\bibnamefont {Mizukami}},
	\bibinfo {author} {\bibfnamefont {Y.}~\bibnamefont {Arai}}, \bibinfo {author}
	{\bibfnamefont {Y.}~\bibnamefont {Sugimura}}, \bibinfo {author}
	{\bibfnamefont {N.}~\bibnamefont {Maejima}}, \bibinfo {author} {\bibfnamefont
		{A.}~\bibnamefont {Machida}}, \bibinfo {author} {\bibfnamefont
		{T.}~\bibnamefont {Watanuki}}, \bibinfo {author} {\bibfnamefont
		{T.}~\bibnamefont {Fukuda}}, \bibinfo {author} {\bibfnamefont
		{T.}~\bibnamefont {Yajima}}, \bibinfo {author} {\bibfnamefont
		{Z.}~\bibnamefont {Hiroi}}, \bibinfo {author} {\bibfnamefont {K.~Y.}\
		\bibnamefont {Yip}}, \bibinfo {author} {\bibfnamefont {Y.~C.}\ \bibnamefont
		{Chan}}, \bibinfo {author} {\bibfnamefont {Q.}~\bibnamefont {Niu}}, \bibinfo
	{author} {\bibfnamefont {S.}~\bibnamefont {Hosoi}}, \bibinfo {author}
	{\bibfnamefont {K.}~\bibnamefont {Ishida}, \bibfnamefont {K.~andMukasa}},
	\bibinfo {author} {\bibfnamefont {T.}~\bibnamefont {Watashige}}, \bibinfo
	{author} {\bibfnamefont {S.}~\bibnamefont {Kasahara}}, \bibinfo {author}
	{\bibfnamefont {J.-G.}\ \bibnamefont {Cheng}}, \bibinfo {author}
	{\bibfnamefont {S.~K.}\ \bibnamefont {Goh}}, \bibinfo {author} {\bibfnamefont
		{Y.}~\bibnamefont {Matsuda}}, \bibinfo {author} {\bibfnamefont
		{Y.}~\bibnamefont {Uwatoko}}, \ and\ \bibinfo {author} {\bibfnamefont
		{T.}~\bibnamefont {Shibauchi}},\ }\href {https://arxiv.org/abs/1704.02057}
{\bibfield  {journal} {\bibinfo  {journal} {arXiv:1704.02057}\ } (\bibinfo
	{year} {2017})}\BibitemShut {NoStop}%
\end{thebibliography}

%merlin.mbs apsrev4-1.bst 2010-07-25 4.21a (PWD, AO, DPC) hacked
%Control: key (0)
%Control: author (72) initials jnrlst
%Control: editor formatted (1) identically to author
%Control: production of article title (-1) disabled
%Control: page (0) single
%Control: year (1) truncated
%Control: production of eprint (0) enabled
%

\end{document}